\documentclass[twocolumn, aps, prd,preprintnumbers]{revtex4-1}

\usepackage[utf8]{inputenc}
\usepackage{braket}
\usepackage{amsmath}
\usepackage{color}
\usepackage{graphicx}
\usepackage{epstopdf}
\usepackage{subcaption}
\usepackage{caption}
\usepackage{amssymb}
\usepackage{amsfonts}
\usepackage{tikz}
\usepackage{hyperref}
\usepackage[format=plain,justification=centerlast,singlelinecheck=true]{caption}
\usepackage[export]{adjustbox}

\hypersetup{
colorlinks,
citecolor=blue,
filecolor=blue,
linkcolor=blue,
urlcolor=blue, 
pdfborder = {0 0 0}, 
linktocpage
}

\renewcommand{\Re}{\text{Re}} 
\renewcommand{\Im}{\text{Im}}

\begin{document}


\preprint{LMU-ASC 03/17}


\title{Effect of acceleration on localized fermionic Gaussian states: \\ From vacuum entanglement to maximally entangled states}


\author{Benedikt Richter$ ^{a,b,c}$}
\author{Krzysztof Lorek$ ^d$}
\author{Andrzej Dragan$ ^d$}
\author{Yasser Omar$ ^{a,b}$}

\affiliation{$^a$Instituto de Telecomunica\c{c}\~oes, Physics of Information and Quantum Technologies Group, Portugal }
\affiliation{$^b$Instituto Superior T\'{e}cnico, Universidade de Lisboa, Portugal}
\affiliation{$^c$Arnold Sommerfeld Center for Theoretical Physics, Department f\"ur Physik, Ludwig-Maximilians-Universit\"at M\"unchen, Germany}
\affiliation{$^d$Institute of Theoretical Physics, University of Warsaw, Poland}


\date{April 4, 2017}

\begin{abstract}

We study the effects of acceleration on fermionic Gaussian states of localized modes of a Dirac field. We consider two wavepackets in a Gaussian state and transform these to an accelerated frame of reference. In particular, we formulate the action of this transformation as a fermionic quantum channel. Having developed the general framework for fermions, we then investigate the entanglement of the vacuum, as well as the entanglement in Bell states. We find that  with increasing acceleration vacuum entanglement increases, while the entanglement of Bell states decreases. Notably, our results have an immediate operational meaning given the localization of the modes.

\end{abstract}


\maketitle

\section{Introduction}

In general spacetimes the notion of particles loses its invariant meaning. One consequence of this observation is the detection of thermal particles in the Minkowski vacuum by a uniformly accelerated observer. This phenomenon is known as the Unruh effect \cite{Unruh:1976db}. It has immediate consequences for quantum states observed by accelerated observers. Their description becomes observer-dependent. In particular, the entanglement present in quantum states cannot be assumed to be independent of the observer \cite{Peres:2002wx, FuentesSchuller:2004xp}.

In the past, several investigations of entanglement in uniformly accelerating systems used nonlocalized Fock states, see e.g. \cite{FuentesSchuller:2004xp, Bruschi:2010mc, Richter:2015wha} and references therein. Therefore, it was not clear how to obtain an operational meaning of entanglement in these states. Furthermore, as pointed out in \cite{Dragan:2012hy, Ahmadi:2016fbd}, it is questionable that these simplified models capture the consequences of acceleration correctly. Another approach that is commonly used to study entanglement in the Minkowski vacuum is employing accelerated Unruh-de Witt detectors \cite{birrell1984quantum}. In these models, a two-level system is linearly coupled to a scalar field and the effect of finite time interactions is studied \cite{benatti2004entanglement}. In particular, it was shown that entanglement can be extracted from vacuum correlations \cite{Salton:2014jaa, Lorek:2014dwa}. The process of extracting entanglement from the vacuum sometimes is referred to as ``entanglement harvesting'' \cite{MartinMartinez:2012sg}. Since these effects are typically small, a direct observation is a challenging task. However, analog systems to study these and related phenomena were proposed \cite{delRey:2011jt, garcia2016entanglement, richter2017}. Another approach to localize quantum states was taken in \cite{Dragan:2012hy, Ahmadi:2016fbd, Kubicki:2016ngz, Dragan:2013dna}. Instead of making use of Unruh-de Witt detectors, the modes constituting the state itself were constructed to be localized. That is also the approach we are following in this work. 

The goal of this work is to give a fully analytic treatment of the effect of acceleration on general localized fermionic Gaussian states. These states are of great relevance, as the class of fermionic Gaussian states contains a broad variety of states such as vacuum states of quadratic Hamiltonians, thermal states and Bell states. Our approach is based on a method that was developed to study bosonic two-mode Gaussian states of localized wave packets \cite{Ahmadi:2016fbd}. Similarly to the bosonic case, we are able to formulate the transformation connecting inertial and accelerated observers as the action of a Gaussian quantum channel.

The framework developed in this work is then used to study the entanglement of the vacuum. In particular, we quantify the amount of entanglement that could possibly be harvested from the vacuum by local detectors and show that it increases with increasing acceleration. Furthermore, we are able to obtain the entanglement in Bell states, a problem that attracted a lot of attention \cite{FuentesSchuller:2004xp, Bruschi:2010mc, Richter:2015wha}, and show that this effect is due to an inevitable mismatch of the wave functions in Minkowski and Rindler spacetime.

The structure of this work is the following. In Sec.\ \ref{framework} we introduce the setting that we study in this work. In Sec.\ \ref{fermionicnoninertial} we calculate the transformation of general fermionic Gaussian states for the case of two uniformly accelerated observers with arbitrary accelerations. With these results in hand, we study the vacuum entanglement in Sec.\ \ref{vacuumcase}. In particular, we compare our findings to the case of bosonic Gaussian states. In Sec.\ \ref{bellstates}, we study the entanglement in Bell states. Finally, we give the conclusions in Sec.\ \ref{conclusions}.

\section{Framework}\label{framework}

\subsection{Outline}

As anticipated above, we study the effect of acceleration on spatially localized fermionic modes. These modes are solutions of the equation of motion for the Dirac field $\hat{\Psi}$. The Dirac equation is obtained by varying the action
\begin{equation}\label{actiondirac}
S_D=\int \text{d}^2 x \sqrt{-g}\left(i\bar{\hat{\Psi}}g_{\mu\nu}\gamma^\mu \partial^\nu \hat{\Psi}-m \bar{\hat{\Psi}}\hat{\Psi}\right),
\end{equation}
where $g$ denotes the determinant of the metric $g_{\mu\nu}$, $\gamma^\mu$ are the gamma matrices, $m$ is the mass of the field and $\bar{\hat{\Psi}}=\hat{\Psi}^\dagger \gamma^0$. The field can be expanded in a complete set of solutions of the Dirac equation.  Since we are interested in modes that are sufficiently localized, we choose to expand $\hat{\Psi}$ in terms of localized wavepackets $\phi_k^\pm$. This gives the expansion
\begin{equation}\label{expansionminkowski}
\hat{\Psi}=\sum_k \left(\phi_k^+ \hat{f}_k +\phi_k^- \hat{g}_k^\dagger \right)
\end{equation}
in Minkowski space, where the $\hat{f}_k^\dagger/\hat{g}_k^\dagger$ are the creation operators for the wavepackets $\phi_k^\pm$ of particles/antiparticles. The Dirac equation dictates the anticommutativity of the creation and annihilation operators; $\{\hat{f}_k , \hat{f}_l\}= 0$, $\{\hat{f}_k^\dagger , \hat{f}_l^\dagger\}= 0$  and $\{\hat{f}_k,\hat{f}_l^\dagger\}= \delta_{kl}$, where the same holds for $\hat{g}_k$ and $\hat{g}_k^\dagger$ with vanishing mixed anticommutators. Alternatively, we can carry out an equivalent expansion in Rindler space and obtain in terms of the respective creation and annihilation operators $\hat{d}_k^\dagger/\hat{e}_k^\dagger$ and $\hat{d}_k/\hat{e}_k$
\begin{equation}\label{expansionrindler}
\hat{\Psi}= \sum_k \left(\psi_k^+ \hat{d}_k +\psi_k^- \hat{e}_k^\dagger\right).
\end{equation}
Here the $\psi_k^\pm$ denote the respective wavepackets and the $\hat{d}_k$ and $\hat{e}_k$ satisfy the same algebra as the $\hat{f}_k$ and $\hat{g}_k$.

Let Alice and Bob be two observers that are each in possession of a (localized) mode of the fermion field $\hat{\Psi}$. These modes have negligible overlap and we denote these by $\phi_\text{I}^+$ and $\phi_\text{II}^+$, respectively. It follows that the corresponding operators commute; $[\hat{f}_\text{I}, \hat{f}_\text{II}^\dagger]=0$. Initially, these modes are prepared in a fermionic Gaussian state $\rho$ that is completely characterized by its first and second moments (covariance matrix). Then to describe the state of the shared pair of modes from an accelerated perspective,  Rindler space offers a convenient reference frame. The modes in both spacetimes are related to each other by a Bogolyubov transformation. Hence, the Gaussianity of state $\rho$ is preserved under such a transformation. The resulting state of Alice's and Bob's modes is thus again a Gaussian of the transformed modes $\psi_\text{I}^\pm$ and $\psi_\text{II}^\pm$. As for bosons \cite{Ahmadi:2016fbd, PhysRevA.63.032312}, the map transforming the covariance matrix of the inertial modes, into the state of the accelerated modes, is a trace preserving Gaussian map $\sigma^{(f)}\rightarrow\sigma^{(d)}$, which takes the form
\begin{equation}\label{trafocovmatrix}
\sigma^{(d)}=M\sigma^{(f)} M^T +N,
\end{equation}
where $M$ and $N$ are $8\times 8$ matrices \cite{Bravyi05} and  we use the superscripts $(f)$ and $(d)$ to indicate the modes that are used to calculate the covariance matrix. One of the main results of the present work is the exact analytic calculation of the matrices $M$ and $N$ for the generic case of two uniformly accelerated observers.

\subsection{Fermionic Gaussian states}\label{fergaustates}

While bosonic Gaussian states are intensively studied in the context of quantum information and extensive literature exists \cite{RevModPhys.77.513, RevModPhys.84.621}, fermionic Gaussian states are much less investigated. Therefore, we insert a brief discussion of these states. The most prominent members of the family of fermionic Gaussian states are vacuum states of quadratic Hamiltonians, thermal states and the Bell states. Even so fermionic Gaussian states share some similarities with their bosonic counterparts, there are crucial differences due to the anticommutativity of fermions. Starting from the creation and annihilation operators $\hat{f}_k$ and $\hat{f}_k^\dagger$ that satisfy the CAR algebra, i.e, $\{\hat{f}_k , \hat{f}_l\}= 0$ and $\{\hat{f}_k,\hat{f}_l^\dagger\}= \delta_{kl}$, we define the Majorana fermion operators  $\hat{c}_k$, as
\begin{equation}\label{defofmajorana}
\hat{c}_{2j-1}=\frac{1}{\sqrt{2}}(\hat{f}_j^\dagger+\hat{f}_j), \hspace{5mm} \hat{c}_{2j}=\frac{1}{i\sqrt{2}} (\hat{f}_j^\dagger-\hat{f}_j).
\end{equation}
These form a Clifford algebra 
\begin{equation}
\{\hat{c}_k,\hat{c}_l\}= \delta_{kl}.
\end{equation}
By definition, the density matrix of an even fermionic Gaussian state can be written as
\begin{equation}
\rho=C \,e^{-\frac{i}{2}\hat{c}^T A \hat{c}},
\end{equation}
where $A$ is a real, antisymmetric matrix and $C$ is the normalization \cite{Bravyi05}. That is, these states are thermal states of a quadratic Hamiltonian $H=\frac{i}{2}\hat{c}^T A \hat{c}$. Upon an $SO(2n)$ transformation $O$ with $n$ being the number of Majorana fermions $\hat{c}_i$, we can write the density matrix in the form
\begin{equation}
\rho=\frac{1}{2^n} \prod_{k=1}^n \left(1+i\lambda_k \hat{c}_{2k-1}' \hat{c}_{2k}'\right),
\end{equation}
where the $\lambda_i$ live in $[-1,1]$ and $\hat{c}'=O^T\hat{c}$. Further, using the Bloch-Messiah reduction \cite{bloch1962canonical}, we can find a basis of modes ${\hat{h}_k}$  such that every pure fermionic Gaussian state can be written in the form 
\begin{equation}
\rho=\prod_{k=1}^n \left(u_k+v_k \hat{h}_{k}^\dagger \hat{h}_{-k}^\dagger\right),
\end{equation}
where $|u_k|^2+|v_k|^2=1$. Even Gaussian states are completely characterized by the corresponding covariance matrix $\sigma_{kl}$ \cite{Bravyi05}.  The real antisymmetric covariance matrix $\sigma_{kl}$ for a Gaussian state $\rho$ is given by
\begin{equation}\label{covmatrix}
\sigma_{kl}= i \,Tr(\rho[\hat{c}_k,\hat{c}_l])= i \langle \hat{c}_k \hat{c}_l - \hat{c}_l \hat{c}_k \rangle.
\end{equation}
The diagonal elements in (\ref{covmatrix}) are vanishing due to the commutator and the covariance matrix is completely characterized by the elements
\begin{equation}
\sigma_{kl}= 2i \langle \hat{c}_k \hat{c}_l \rangle,\hspace{10mm}\text{with} \,\, k>l.
\end{equation}
All higher moments of $\rho$ can be obtained by Wick's theorem \cite{Bravyi05}. So far we concentrated on even states. While these are completely characterized by $\sigma_{kl}$, the description of odd Gaussian states requires, in general, also the knowledge of the first moments
\begin{equation}\label{firstmoments}
 Tr(\rho \hat{c}_k)=  \langle \hat{c}_k \rangle.
\end{equation}
These are naturally vanishing for even states due to the vanishing commutator with the parity operator $P$, $[P, \rho]=0$ \cite{Bravyi05}. More information regarding fermionic Gaussian states and Gaussian linear maps can, for example, be found in \cite{kraus2009pairing, de2013power, bravyi2005classical,  Bradler:2010st, greplova2016degradability}.

\subsection{Dirac field in Rindler spacetime}

We briefly introduce solutions of the Dirac equation that is obtained by varying action (\ref{actiondirac}) with respect to $\bar{\hat{\Psi}}$. For simplicity, we restrict our attention to the $1+1$ dimensional case. However, the generalization to $3+1$ dimensions is straightforward. Furthermore, we use units such that $c=\hbar=1$ throughout the entire work. We start by deriving the solutions of the Dirac equation in Minkowski coordinates $(t,x)$. The Dirac equation takes the form 
\begin{equation}
i\partial_t\hat{\Psi}=\left(-i\alpha_3\partial_x+m\beta\right)\hat{\Psi},
\end{equation}
where $m$ denotes the mass of the field and the matrices $\beta$ and $\alpha_3$ are given by
\begin{equation}\label{matrices}
\beta=\left( \begin{array}{cc}
1 & 0  \\
0& -1   \end{array} \right), \hspace{10mm} \alpha_3=\left( \begin{array}{cc}
0 & 1  \\
1& 0   \end{array}\right).
\end{equation} 
We find the solutions in Minkowski coordinates to be given by
\begin{align}
u_{k,\pm}=&\frac{1}{\sqrt{4\pi\omega_k}}  \left(
\begin{array}{c}
\sqrt{\omega_k\pm m}\\
\pm\sqrt{\omega_k\mp m}
\end{array}
\right)e^{\mp i\omega_k t+ikx}  ,
\end{align} 
where $\omega_k=\sqrt{k^2+m^2}$. These are normalized as given in (\ref{orthonormalization}). For further details, see \cite{birrell1984quantum}. Next, to describe the physics of an accelerated observer \cite{Crispino:2007eb, takagi1986vacuum}, we work in Rindler coordinates 
\begin{subequations}\label{coordtrafo}
\begin{align}
t=& \chi \sinh(a\eta),\\
x=&\chi \cosh(a\eta),
\end{align} 
\end{subequations}
 where $a$ is a parameter that does not have an immediate physical meaning. However, if we consider the worldline of a particle with a proper acceleration $\cal A$, namely $\chi=\frac{1}{\cal A}$, its proper time $\tau$ can be related
to the coordinate time $\eta$ as ${\cal A} \tau =a \eta$. In Rindler coordinates, the equation of motion obtained from (\ref{actiondirac}) is given by
\begin{equation}\label{diracinrindler}
i \frac{1}{a}\partial_\eta\hat{\Psi}=\left(m\chi \beta-\frac{i}{2} \alpha_3-i\chi\alpha_3\partial_{\chi}\right)\hat{\Psi},
\end{equation}
where $m$ denotes the mass of the field and the matrices $\beta$ and $\alpha_3$ are given by (\ref{matrices}). Solving the Dirac equation using the ansatz $w^\pm_{\text{I}\Omega}(\eta,\chi)= e^{\mp i\Omega\eta}w^\pm(\chi)$ for particles and antiparticles respectively, one finds the normalized solutions to be given by
\begin{widetext}
\begin{align}
w^\pm_{\text{I}\Omega}=&\sqrt{\frac{m \cosh(\frac{\pi \Omega}{a})}{2\pi^2 a}} \left(
\begin{array}{c}
K_{\pm i\frac{\Omega}{a}+\frac{1}{2}}(m\chi)+i K_{\pm i\frac{\Omega}{a}-\frac{1}{2}}(m\chi)\\
- K_{\pm i\frac{\Omega}{a}+\frac{1}{2}}(m\chi)+i K_{\pm i\frac{\Omega}{a}-\frac{1}{2}}(m\chi)
\end{array}
\right)  e^{\mp i\Omega\eta}, \hspace{13mm} \text{in Rindler wedge $I$},\\
w^\pm_{\text{II}\Omega}=&\sqrt{\frac{m \cosh(\frac{\pi \Omega}{a})}{2\pi^2 a}}  \left(
\begin{array}{c}
K_{\pm i\frac{\Omega}{a}+\frac{1}{2}}(-m\chi)+i K_{\pm i\frac{\Omega}{a}-\frac{1}{2}}(-m\chi)\\
- K_{\pm i\frac{\Omega}{a}+\frac{1}{2}}(-m\chi)+i K_{\pm i\frac{\Omega}{a}-\frac{1}{2}}(-m\chi)
\end{array}
\right) e^{\pm i\Omega\eta}, \hspace{6mm} \text{  in Rindler wedge $II$},
\end{align}
\end{widetext}
where $K_{ y}(x)$ is the modified Bessel function of the second kind of order $y$.  The Dirac inner product of the mode functions $\omega_1$ and $\omega_2$ is defined as $\left(\omega_1(x),\omega_2(x)\right)_\Sigma=\int \text{d}\Sigma^\mu \omega_1^\dagger(x)\gamma^0\gamma_\mu \omega_2(x)$, where $\text{d}\Sigma^\mu=n^\mu \text{d}s$ is space-like with normal vector $n^\mu$ and volume element $\text{d}s$. It has the properties $\left(\omega_1(x),\omega_2(x)\right)^*_\Sigma=\left(\omega_1^*(x),\omega_2^*(x)\right)_\Sigma =\left(\omega_2(x),\omega_1(x)\right)_\Sigma$. Here, in Rindler coordinates, we make the particular choice
\begin{equation}
\left(\omega_1,\omega_2\right)=\int \text{d}\chi\, \omega_1^\dagger\omega_2,
\end{equation}
where we suppressed the coordinate dependence of the mode functions. The positive and negative energy solutions satisfy the relations
\begin{equation}\label{orthonormalization}
(w^\pm_{\text{I}\Omega},w^\pm_{\text{I}\Omega'})= \delta( \Omega-\Omega'),\hspace{5mm}(w^\pm_{\text{I}\Omega},w^\mp_{\text{I}\Omega'})=0.
\end{equation}
The decomposition of $\hat{\Psi}$ in terms of localized modes is given by
\begin{equation}\label{fieldexpansion}
\hat{\Psi}=\sum_k \left(\phi_k^+ \hat{f}_k +\phi_k^- \hat{g}_k^\dagger \right)= \sum_k \left(\psi_k^+ \hat{d}_k +\psi_k^- \hat{e}_k^\dagger\right),
\end{equation}
where $\hat{f}_k/\hat{g}_k$ are the annihilation operators of particles/antiparticles in the Minkowski case and $\hat{d}_k/\hat{e}_k$ are the annihilation operators of particles/antiparticles in the Rindler case. These are related to the Rindler particle and antiparticle operators, $\hat{b}_\Omega$ and $\hat{a}_\Omega$,  by 
\begin{subequations}\label{relationdb}
\begin{align}
\hat{d}_\text{I}=&\int \text{d}\Omega (\psi_\text{I}^+, w^+_{\text{I}\Omega})\hat{b}_{\text{I}\Omega},\\
\hat{d}_\text{II}=&\int \text{d}\Omega (\psi_\text{II}^+, w^+_{\text{II}\Omega})\hat{b}_{\text{II}\Omega},\\
\hat{e}_\text{I}^\dagger=&\int \text{d}\Omega (\psi_\text{I}^-, w^-_{\text{I}\Omega})\hat{a}^{\dagger}_{\text{I}\Omega},\\
\hat{e}_\text{II}^\dagger=&\int \text{d}\Omega (\psi_\text{II}^-, w^-_{\text{II}\Omega})\hat{a}^{\dagger}_{\text{II}\Omega}.
\end{align}
\end{subequations}
The Minkowski vacuum $|0\rangle_\text{M}$ is related to the Rindler vacuum $|0\rangle_\text{R}$ by a squeezing operator $S$ as
\begin{equation}
|0\rangle_\text{M}=S |0\rangle_\text{R},
\end{equation}
where $S$ acts on $\hat{b}_{\text{I}\Omega}$ and $\hat{a}_{\text{II}\Omega}$  as
\begin{subequations}\label{squeezing}
\begin{align}
S^\dagger \hat{b}_{\text{I}\Omega} S=& \cos(r_\Omega) \hat{b}_{\text{I}\Omega}-\sin(r_\Omega) \hat{a}_{\text{II}\Omega}^{\dagger},\\
S^\dagger \hat{a}_{\text{II}\Omega} S=& \cos(r_\Omega) \hat{a}_{\text{II}\Omega}+\sin(r_\Omega) \hat{b}_{\text{I}\Omega}^{\dagger}
\end{align}
\end{subequations}
with $\tan(r_\Omega)=e^{-\frac{\pi |\Omega|}{a}}$. To obtain the transformations of the operators $\hat{b}_{\text{II}\Omega}$ and $\hat{a}_{\text{I}\Omega}$, one simply has to interchange $\hat{a}\leftrightarrow \hat{b}$ in (\ref{squeezing}).

\subsection{Modes}\label{modes}
  
  \begin{figure}[t]

\includegraphics[scale=1]{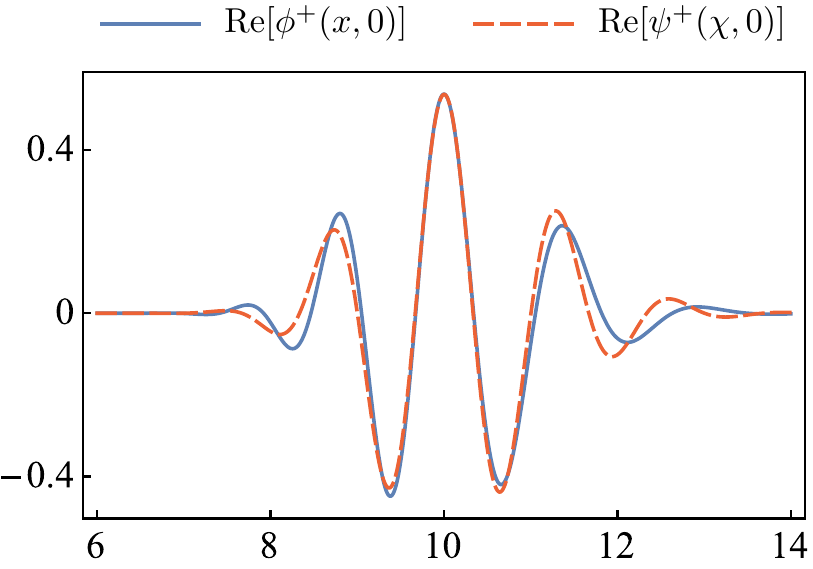}
\caption{ Comparison between the real parts of the spatial modes $\phi^+$ and $\psi^+$ for the following choice of parameters: $x_0^{-1} = {\cal A}=0.1$, $L=2$, $\Omega_0=4.71$, $m=0.1$. These modes are localized and, therefore, a single proper acceleration ${\cal A}$ can be assigned.}
\label{picmodes}
\end{figure}

Most of our calculations are general and do not depend on the particular choice of modes. However, when considering concrete examples, namely in Sec.\ \ref{vacuumcase} and  Sec.\ \ref{bellstates}, we have to make a specific choice. Physically, cavity modes are suitable candidates for localized  solutions. In particular, motivated by the solutions obtained in \cite{Friis:2013eva}, for $\Lambda\in \{\text{I},\text{II}\}$, we consider the Minkowski modes
\begin{equation}\label{inercavity}
\phi_\Lambda^+(x,0)= C_\phi \, \xi_\phi(k) \, e^{-2\left(\frac{x_0}{L}\log(\frac{x}{x_0})\right)^2+ik x},
\end{equation}
where   $L$ is the width of the wave packet that is centred at $x_0$ and $\xi_\phi(k)$ is 
\begin{equation}
\xi_\phi(k)=\left(
\begin{array}{c}
\cos\left(\frac{\kappa}{2}\right)+\sin\left(\frac{\kappa}{2}\right)\\
\cos\left(\frac{\kappa}{2}\right)-\sin\left(\frac{\kappa}{2}\right)
\end{array}
\right)  e^{-\frac{i}{2}\kappa}e^{-i k x_0}
\end{equation}
with $\kappa=\arctan(\frac{m}{k})$ and $C_\phi$ being a constant of normalization. As in \cite{Ahmadi:2016fbd}, we use a Gaussian profile to sufficiently localize the mode.

For the accelerated case, we choose wave packets of particle solutions of the Dirac equation in Rindler space, (\ref{diracinrindler}). Specifically, the localized modes we work with are 
\begin{equation}\label{acccavity}
 \psi_\Lambda^+(\chi,0)= C_\psi \, \xi_\psi(\chi) \, e^{-2\left(\frac{\chi_0}{L}\log(\frac{\chi}{\chi_0})\right)^2},
\end{equation}
where  $\xi_\psi(\chi)$ is given by the spinor
\begin{equation}
\xi_\psi(\chi)=\left(
\begin{array}{c}
I_{i\frac{ \Omega }{a}-\frac{1}{2}}(m \chi )+i I_{i\frac{\Omega }{a}+\frac{1}{2}}(m \chi )\\
I_{i\frac{ \Omega }{a}-\frac{1}{2}}(m \chi )-i I_{i\frac{ \Omega }{a}+\frac{1}{2}}(m \chi )
\end{array}
\right)
\end{equation}
and $C_\psi$ is 
\begin{equation}
C_\psi=\tilde{C}_\psi\left( I_{-i\frac{ \Omega }{a}-\frac{1}{2}}\left(m\chi_0\right)-I_{-i\frac{ \Omega }{a}+\frac{1}{2}}\left(m\chi_0\right)\right)
\end{equation}
with the constant of normalization, $\tilde{C}_\psi$ and $I_{i\mu}(x)$ being the modified Bessel functions of the first kind. The modes are plotted in Fig.\ \ref{picmodes}. 
 
The physical motivation for choosing localized cavity modes is that these modes can possibly be detected by a local detector (in our case a cavity). Therefore, the choice of localized modes enables us to assign an operational meaning to quantities like entanglement. Furthermore, the modes (\ref{inercavity}) and (\ref{acccavity}) are purely positive frequency solutions. This guarantees, that the average particle and antiparticle numbers are zero in the Rindler vacuum, as expected. In consequence, the modes do necessarily have noncompact support. However, due to the Gaussian profile they are still sufficiently localized.

\section{Accelerated fermionic Gaussian states}\label{fermionicnoninertial}

In this section, we study the Bogolyubov transformation of fermionic Gaussian states  relating modes in Minkowski space to those of Rindler space. We give the analytical solution for the general case of two uniformly accelerated observers.

\subsection{Transformation of the Minkowski vacuum}

To study the effect of acceleration on the Minkowski vacuum, we restrict our attention to two localized fermion modes with respective creation operators $\hat{f}_1^\dagger$ and $\hat{f}_2^\dagger$. Then, using (\ref{defofmajorana}), we can  write the corresponding Majorana operators $\hat{c}_i$  as
\begin{equation}
\left(
\begin{array}{c}
\hat{c}_1\\
\hat{c}_2\\
\hat{c}_3\\
\hat{c}_4\\
\end{array}
\right)
=\frac{1}{\sqrt{2}}
\left(
\begin{array}{c}
\hat{f}_\text{I}^\dagger+\hat{f}_\text{I}\\
\frac{1}{i} (\hat{f}_\text{I}^\dagger-\hat{f}_\text{I})\\
\hat{f}_\text{II}^\dagger+\hat{f}_\text{II}\\
\frac{1}{i} (\hat{f}_\text{II}^\dagger-\hat{f}_\text{II})\\
\end{array}
\right).
\end{equation}
From (\ref{covmatrix}) it is clear that in the case of two modes, there are six independent entries in the fermionic covariance matrix. The same is true for the corresponding antiparticles. Therefore, we find the covariance matrix  of the Minkowski vacuum 
\begin{align}\label{covmatrixinertial}
\sigma^{(f)}_\text{M}=&\left(\begin{array}{cccccccc}
0 & 1& 0 & 0  & 0 & 0& 0 & 0     \\
-1 & 0& 0 & 0  & 0 & 0& 0 & 0    \\
0 & 0& 0 & 1 & 0 & 0& 0 & 0     \\
0 & 0 & -1 & 0 & 0 & 0& 0 & 0 \\
0 & 0 & 0 & 0 & 0 & 1& 0 & 0 \\
0 & 0 & 0 & 0 & -1 & 0& 0 & 0 \\
0 & 0 & 0 & 0 & 0 & 0& 0 &1 \\
0 & 0 & 0 & 0 & 0 & 0& -1 & 0 \\
\end{array}\right),
\end{align}
where this expression describes two particle modes and two antiparticle modes. However, it generalizes trivially to an arbitrary number of particles and antiparticle modes; see Appendix \ref{appminkvac} for details. As it is known for pure states \cite{Bravyi05}, $\sigma^{(f)^T}_\text{M} \sigma^{(f)}_\text{M}=1$ and the first moments vanish, as the vacuum is an even state.  To obtain  the covariance matrix $\sigma^{(d)}_\text{M}$ of the Minkowski vacuum in the Rindler frame, we need to transform the operators $\hat{f}$ and $\hat{g}$ (for details see Appendix \ref{appminkvactrafo}). We find the covariance matrix of the Minkowski vacuum in the Rindler basis to be of the form
\begin{equation}\label{covmatrixnoninertial}
\sigma^{(d)}_{}=\left(\begin{array}{cc}
\sigma_{+}^{(d)} & \sigma_{c}^{(d)}    \\
\tilde{\sigma}_{c}^{(d)}  & \sigma_{-}^{(d)}  
\end{array}\right),
\end{equation}
where the block $\sigma_{+}^{(d)}$ captures the effect of acceleration on the respective mode and displays the thermal character of the Unruh effect ($\sigma_{-}^{(d)}$ is the analog for antiparticles). $\sigma_{c}^{(d)} $ and $\tilde{\sigma}_{c}^{(d)}$ describe the correlations between particles and antiparticles that arise due to the acceleration. Explicitly, (\ref{covmatrixnoninertial}) reads
\begin{widetext}
\begin{equation}\label{covmatrixnoninertialdntriv}
\sigma^{(d)}_{}=\left(\begin{array}{cccccccc}
0 & N_\text{I}^+& 0 & 0   & 0 & 0& \Im[N_\text{I,II}^+] & \Re[N_\text{I,II}^+]   \\
-N_\text{I}^+ & 0& 0 & 0   & 0 & 0& \Re[N_\text{I,II}^+] & -\Im[N_\text{I,II}^+]  \\
0 & 0 & 0 & N_\text{II}^+  &   -\Im[N_\text{I,II}^-] & -\Re[N_\text{I,II}^-]   & 0 & 0  \\
0& 0& -N_\text{II}^+ & 0  &  -\Re[N_\text{I,II}^-] & \Im[N_\text{I,II}^-]  & 0 & 0 \\
0 & 0& \Im[N_\text{I,II}^-] & \Re[N_\text{I,II}^-]  &  0 & N_\text{I}^-& 0 & 0  \\
0 & 0& \Re[N_\text{I,II}^-] & -\Im[N_\text{I,II}^-]  &  -N_\text{I}^- & 0& 0 & 0 \\
-\Im[N_\text{I,II}^+] & -\Re[N_\text{I,II}^+]   & 0 & 0  &  0 & 0 & 0 & N_\text{II}^- \\
-\Re[N_\text{I,II}^+] & \Im[N_\text{I,II}^+]  & 0 & 0& 0& 0& -N_\text{II}^- & 0 
\end{array}\right),
\end{equation}
\end{widetext}
where the entries of the matrix are given by
\begin{subequations}\label{covmatelefermi}
\begin{align}
N_\text{I}^\pm=& 1-2\int \text{d}\Omega\, \frac{|(\psi_\text{I}^\pm , w^\pm_{\text{I}\Omega})|^2}{1+e^{\frac{2\pi\Omega}{a}}} \, ,\label{diagonalelement1}\\
N_\text{II}^\pm=& 1-2\int \text{d}\Omega\, \frac{|(\psi_\text{II}^\pm , w^\pm_{\text{II}\Omega})|^2}{1+e^{\frac{2\pi\Omega}{a}}}\,,\label{diagonalelement2}\\
N_\text{I,II}^\pm=&-2 \int \text{d}\Omega\, \frac{(\psi_\text{I}^\pm, w^\pm_{\text{I}\Omega}) (\psi_\text{II}^\mp, w^\mp_{ \text{II}\Omega})}{1+e^{\frac{2\pi\Omega}{a}}}\,e^{\frac{\pi\Omega}{a}}\label{diagonalelement3}
\end{align}
\end{subequations}
and we used $(\psi_\text{I}^\pm, w^\pm_{\text{I} \Omega})^*=(\psi_\text{I}^\mp, w^\mp_{\text{I} \Omega})$. We note that (\ref{covmatrixnoninertial}) reduces to  (\ref{covmatrixinertial}) in the limit of vanishing accelerations, $a\to 0$. That is, the correlated noise is absent in this limit.

\subsection{Transformation of general fermionic Gaussian states}\label{trafoofgeneralstates}

Using the results of the previous section, we derive the transformation for general fermionic Gaussian states. We recall that the transformed covariance matrix $\sigma^{(d)}$ can be obtained using (\ref{trafocovmatrix}).  Therefore, we can completely describe the transformation of general Gaussian states by the two matrices $M$ and $N$ that we calculate in the following. To obtain $M$, we consider the transformation of the first moments (\ref{firstmoments}). The Bogolyubov transformations can be obtained from (\ref{fieldexpansion}) and read
\begin{subequations}\label{gernaltrafo1}
\begin{align}
\hat{d}_l=&\sum_k (\psi_l^+, \phi_k^+) \hat{f}_k + (\psi_l^+, \phi_k^-) \hat{g}_k^\dagger ,\\
\hat{e}_l^\dagger=&\sum_k (\psi_l^-, \phi_k^+) \hat{f}_k +(\psi_l^-, \phi_k^-)  \hat{g}_k^\dagger  ,
\end{align}
\end{subequations}
where, in the following, we denote the overlaps by $\alpha_\text{I}^+=(\psi_\text{I}^+, \phi_\text{I}^+)$, $\beta_\text{I}^+=-(\psi_\text{I}^+, \phi_\text{I}^-)$, $\alpha_\text{I}^-=(\psi_\text{I}^-, \phi_\text{I}^-)^*$, $\beta_\text{I}^-=-(\psi_\text{I}^-, \phi_\text{I}^+)$, and accordingly for modes $\text{II}$. We note that both the Minkowski wave packets $\phi_{\text{I}/\text{II}}^\pm$ as well as the Rindler wave packets $\psi_{\text{I}/\text{II}}^\pm$ are composed of either only  particles or only  antiparticles. Therefore, the overlaps $\beta_{\text{I}/\text{II}}^\pm$ are vanishing and we can write (\ref{gernaltrafo1}) as
\begin{equation}\label{gernaltrafo2}
\hat{d}_l=\alpha_\text{I}^+ \hat{f}_l,\hspace{10mm} \hat{e}_l=\alpha_\text{I}^-  \hat{g}_l\, .
\end{equation}
The transformation of the first moments can be written as
\begin{equation}
\langle \hat{c}^{(d)}\rangle= M\langle \hat{c}^{(f)}\rangle
\end{equation}
with the block-diagonal $8\times 8$-matrix
\begin{equation}\label{Mmatrix}
M=\left(\begin{array}{cccc}
M(\alpha_\text{I}^+) & 0 &0&0     \\
0 & M(\alpha_\text{II}^+)  &0&0\\
0&0&M(\alpha_\text{I}^-)&0    \\
0&0&0&M(\alpha_\text{II}^-)    
\end{array}\right),
\end{equation}
where we defined $M(\cdot)$ as the matrix given by
\begin{equation}
M(\cdot)=\left(\begin{array}{cc}
\Re[\cdot] & \Im[\cdot]      \\
-\Im[\cdot] & \Re[\cdot]      \\
\end{array}\right).
\end{equation}
We now gathered all quantities that we need to specify the transformation of a general Gaussian state characterized by $\langle \hat{c}^{(f)}\rangle$ and $\sigma^{(f)}$. First, we  explicitly calculate the noise matrix $N$. The matrix describing the noise arising due to acceleration is given by
\begin{equation}\label{noisematrix}
N=\sigma^{(d)}-M \sigma^{(f)}_\text{M} M^T,
\end{equation}
where $\sigma^{(d)}$ is given in (\ref{covmatrixnoninertial}). Therefore, in total we obtain
\begin{equation}\label{explnoise}
N =\left(\begin{array}{cc}
N_{+} & N_{\text{vac}}    \\
\tilde{N}_{\text{vac}}   & N_- 
\end{array}\right),
\end{equation}
where $N_{\text{vac}}$ and $\tilde{N}_{\text{vac}}$ are given by the off-diagonal blocks of (\ref{covmatrixnoninertial}) and
\begin{widetext}
\begin{equation}
N_{\pm}=\left(\begin{array}{cccc}
0 & N_\text{I}^\pm-|\alpha_{\text{I}}^\pm|^2& 0 & 0     \\
-N_\text{I}^\pm+|\alpha_{\text{I}}^\pm|^2 & 0& 0 & 0    \\
0 & 0 & 0 & N_{\text{II}}^\pm-|\alpha_{\text{II}}^\pm|^2     \\
0& 0& -N_{\text{II}}^\pm+|\alpha_{\text{II}}^\pm|^2 & 0
\end{array}\right).
\end{equation}
\end{widetext}
With the matrices $M$ and $N$ we now have all data that is necessary to completely characterize the effect of acceleration to fermionic Gaussian states. In the following, we use the developed formalism to quantify vacuum entanglement and, as a example of a nonvacuum state, we study a maximally entangled state, the Bell state $|B\rangle=\frac{1}{\sqrt{2}}(|00\rangle+|11\rangle)$.

\section{Vacuum entanglement}\label{vacuumcase}

\subsection{Vacuum entanglement for fermions}

In this section, we study the entanglement in the vacuum as seen by an accelerated observer. As a measure of entanglement we employ the logarithmic negativity. In \cite{eisler2015partial}, it was shown that the partial transpose of a fermionic  Gaussian state is, in general, not Gaussian and therefore it is difficult to calculate the negativity exactly. However, it was shown that a lower bound can be obtained \cite{eisert2016entanglement, eisler2015partial}. Building on this construction, we derive the respective bound  $\tilde{\mathcal{E}}_\mathcal{N}$ for our case in Appendix \ref{lognegativity}. It is given by
\begin{align}\label{lowerboundfermiond0}
\tilde{\mathcal{E}}_\mathcal{N}=&\ln(\frac{1}{2}(1+N_\text{I}^+N_\text{II}^-+|N_\text{I,II}^+|^2  +\nonumber\\
+& \Re[\sqrt{(N_\text{I}^+-N_\text{II}^-)^2-4|N_\text{I,II}^+|^2}]+\nonumber\\
+& \Im[\sqrt{(N_\text{I}^+-N_\text{II}^-)^2-4|N_\text{I,II}^+|^2}] ))
\end{align}
where $N_\text{I}^+$, $N_\text{II}^-$ and $N_\text{I,II}^+$ are the elements of the covariance matrix (\ref{covmatrixnoninertial}) that are given by (\ref{diagonalelement1}) - (\ref{diagonalelement3}).

\begin{figure}[t]
\includegraphics[scale=1.05]{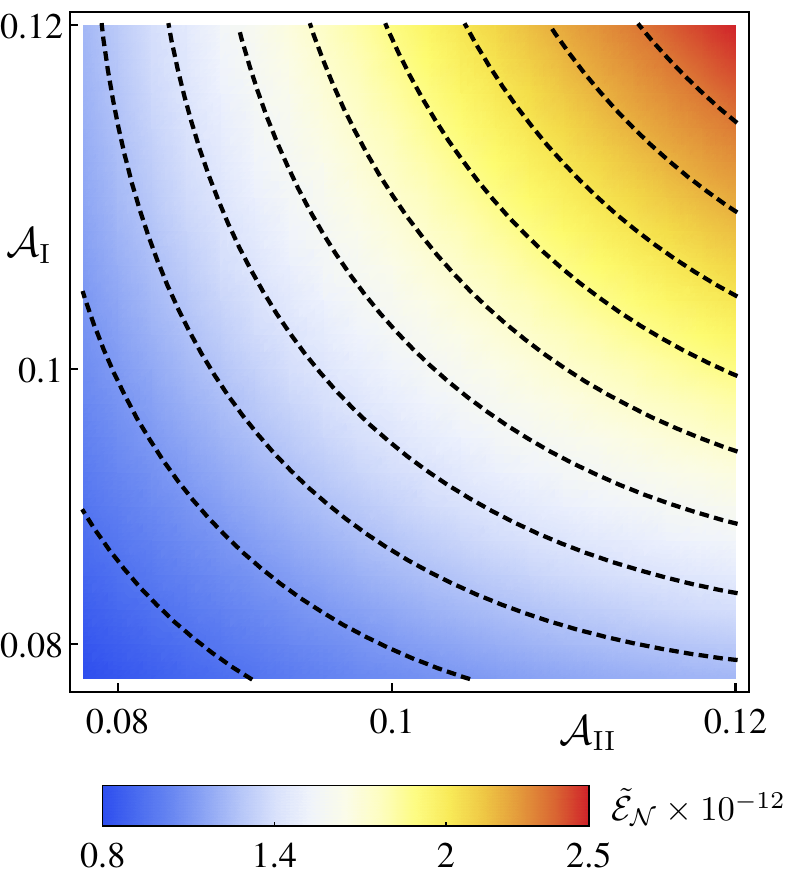}
\caption{Negativity $\tilde{\mathcal{E}}_\mathcal{N}$ of the Minkowski vacuum  as a function of the proper accelerations. The parameters are $m=0.1$, $L=2$ and $\Omega_0=4.71$. The vacuum entanglement increases with increasing acceleration.}
\label{picnegvac}
\end{figure}

To obtain the entanglement of the vacuum, we choose the modes as described in Sec.\ \ref{modes} and solve (\ref{lowerboundfermiond0}) numerically. We show the results for the entanglement between two modes of different accelerations in Fig.\ \ref{picnegvac}.

Firstly, it is important to emphasize that $\tilde{\mathcal{E}}_\mathcal{N}$ is a lower bound for the negativity and therefore it is difficult to make strong quantitative statements. However, it is reasonable to assume that the actual value of the negativity is close to the lower bound $\tilde{\mathcal{E}}_\mathcal{N}$. Therefore, in the following, we refer to $\tilde{\mathcal{E}}_\mathcal{N}$ as negativity, keeping this subtlety in mind.

The entanglement between the state of a mode in Rindler wedge I and its respective counterpart in wedge II increases with increasing acceleration, see Fig.\ \ref{picnegvac}. For the parameter regime we are interested in, the entanglement is of the order $10^{-12}$, as measured by the logarithmic negativity. That is, the particles that are produced due to the Unruh effect are correlated across the acceleration horizon. Interestingly, the vacuum entanglement has an operational meaning, as it can, in principle, be extracted by suitable detectors \cite{Salton:2014jaa, Lorek:2014dwa}.

\subsection{Comparison to bosons}\label{bosons}

The aim of this section is to compare our results for fermionic states to the results in the  case of localized modes of a massive scalar field studied in \cite{Ahmadi:2016fbd}. For bosonic Gaussian states the covariance matrix $\sigma^{(d)}_{\text{bos}}$ of the vacuum takes the form 
\begin{equation}\label{covmatrixnoninertialbosons}
\sigma^{(d)}_{\text{bos}}=
\left(\begin{array}{cccc}
N_\text{I}^b & 0& \Re[N_\text{I,II}^b] & \Im[N_\text{I,II}^b]      \\
0 & N_\text{I}^b& \Im[N_\text{I,II}^b] & -\Re[N_\text{I,II}^b]     \\
\Re[N_\text{I,II}^b] & \Im[N_\text{I,II}^b] & N_\text{II}^b &       \\
\Im[N_\text{I,II}^b] & \-Re[N_\text{I,II}^b] & 0 & N_\text{II}^b
\end{array}\right),
\end{equation}
where the matrix elements are given by
\begin{subequations}\label{covmatbosonele}
\begin{align}
N_{\text{I}/\text{II}}^b=& 1+2\int \text{d}\Omega \frac{|(\psi_\text{I/II}, w_{\text{I}/\text{II} \Omega})|^2 }{e^{\frac{2\pi\Omega}{a}}-1},\\
N_\text{I,II}^b=& 2\int \text{d}\Omega \frac{(\psi_\text{I}, w_{\text{I} \Omega}) (\psi_\text{II}, w_{\text{II} \Omega})}{e^{\frac{2\pi\Omega}{a}}-1}\,e^{\frac{\pi\Omega}{a}},
\end{align}
\end{subequations}
where $\psi_\text{I}$, $\psi_\text{II}$ are localized modes and the $w_{\text{I}/\text{II} \Omega}$ are solutions of the Klein-Gordon equation spanning the Hilbert space for regions I and II, respectively. The effect of acceleration on bosons is manifest in the covariance matrix (\ref{covmatrixnoninertialbosons}). While the Minkowski vacuum is described by the covariance matrix $\sigma^{(\text{bos})}_{M}=\mathbb{I}$, for nonvanishing acceleration, correlations build up and the off-diagonal elements increase.  As expected, the occupations are governed by the Bose-Einstein distribution that are explicit in the matrix elements (\ref{covmatbosonele}). This contrasts with the case of fermions where the occupation is characterized by the Fermi-Dirac distribution, (\ref{covmatelefermi}). Furthermore, due to the presence of antiparticles for fermions, the correlations that are build up in fermionic Gaussian states are between particles and antiparticles, while for bosons particles and antiparticles are identical. 

Besides the differences outlined above, the results for fermions qualitatively agree with the findings for bosons. Quantitatively,  the entanglement we observed for fermions is less than in the bosonic case. In the parameter regime we studied, bosons develop ten times more entanglement \cite{Ahmadi:2016fbd}. However, one should keep in mind that we were calculating a lower bound for the negativity and that the amount of entanglement also depends on the explicit choice of modes.

In the next section, we go beyond vacuum entanglement and study the degradation of entanglement for Bell states.

\section{Entanglement in Bell states}\label{bellstates}

Interestingly, Bell states of fermions are fermionic Gaussian states, while maximally entangled states of two bosons are nonGaussian. This fact, enables us to apply our formalism to Bell states. Therefore, as an example for a nonvacuum state, we consider a maximally entangled state $|B\rangle$ of two particles in the following. We choose
\begin{equation}\label{bellstate}
|B\rangle=\frac{1}{\sqrt{2}}(|00\rangle+|11\rangle)=\frac{1}{\sqrt{2}}(1+\hat{f}_1^\dagger \hat{f}_2^\dagger)|0\rangle.
\end{equation}
This state is an even Gaussian state and it is described by the following covariance matrix (cf.\ Appendix \ref{appminkbell})
\begin{equation}
\sigma^{(f)}_{\text{Bell}}=\left(\begin{array}{cccccccc}
0 & 0&0 &  1 & 0& 0& 0& 0    \\
0 & 0& 1& 0 & 0& 0& 0& 0   \\
0 & - 1 & 0 & 0  & 0& 0& 0& 0  \\
- 1 & 0 & 0 & 0& 0& 0& 0& 0   \\
0 & 0 & 0 & 0& 0& 1& 0& 0   \\
0 & 0 & 0 & 0& -1& 0& 0& 0   \\
0 & 0 & 0 & 0& 0& 0& 0& 1   \\
0 & 0 & 0 & 0& 0& 0& -1& 0   
\end{array}\right).
\end{equation}
The transformed state is obtained according to equation (\ref{trafocovmatrix}) with the matrix $M$ given in (\ref{Mmatrix}) and $N$ given in (\ref{explnoise}). Then the transformed covariance matrix, neglecting some terms much smaller than 1, is given by
\begin{widetext}
\begin{equation}\label{transformedcovmatbell}
\sigma^{(d)}_\text{Bell}=\left(\begin{array}{cccc}
0 & N_\text{I}^+-|\alpha_\text{I}^+|^2 & (\sigma^{(d)}_\text{Bell})_{13} & (\sigma^{(d)}_\text{Bell})_{14}   \\
_-N_\text{I}^++|\alpha_\text{I}^+|^2& 0&   (\sigma^{(d)}_\text{Bell})_{23}   &     (\sigma^{(d)}_\text{Bell})_{24}     \\
- (\sigma^{(d)}_\text{Bell})_{13} & -  (\sigma^{(d)}_\text{Bell})_{23} & 0 & N_\text{II}^+ -|\alpha_\text{II}^+|^2    \\
-(\sigma^{(d)}_\text{Bell})_{14}  & -  (\sigma^{(d)}_\text{Bell})_{24} & -N_\text{II}^++|\alpha_\text{II}^+|^2 & 0
\end{array}\right),
\end{equation}
\end{widetext}
where 
\begin{subequations}
\begin{align}
(\sigma^{(d)}_\text{Bell})_{13}=&\Im[\alpha_\text{II}^+] \Re[\alpha_\text{I}^+] + \Im[\alpha_\text{I}^+] \Re[\alpha_\text{II}^+],\\
(\sigma^{(d)}_\text{Bell})_{14}=& \frac{1}{2} \left(\alpha_\text{I}^+ \alpha_\text{II}^+ + (\alpha_\text{I}^+\alpha_\text{II}^+)^*\right), \\
(\sigma^{(d)}_\text{Bell})_{23}=&\frac{1}{2} \left(\alpha_\text{I}^+ \alpha_\text{II}^+ + (\alpha_\text{I}^+\alpha_\text{II}^+)^*\right),\\
(\sigma^{(d)}_\text{Bell})_{24}=&-\Im[\alpha_\text{II}^+] \Re[\alpha_\text{I}^+] - \Im[\alpha_\text{I}^+] \Re[\alpha_\text{II}^+]
\end{align}
\end{subequations}
with $\alpha_{\text{I}/\text{II}}^+$ as defined in (\ref{gernaltrafo2}). In (\ref{transformedcovmatbell}), we neglected contributions from the vacuum noise matrix $N$, as these are insignificant corrections to $\sigma^{(d)}_\text{Bell}=M\sigma^{(f)}_\text{Bell} M^T$. 

To quantify the entanglement we employ a lower bound for the negativity given by 
\begin{align}\label{lowerboundfermionbell}
\tilde{\mathcal{E}}_\mathcal{N}=\ln(\frac{1}{2}+&\frac{\Re[\nu^+\nu^- +\nu^+-\nu^-]}{2}+\nonumber\\
+&\frac{\Im[\nu^+\nu^-+\nu^+-\nu^-]}{2}),
\end{align}
where the $\nu^\pm$ can be calculated from (\ref{transformedcovmatbell}); cf.\ (\ref{lowerboundneg}) in Appendix \ref{lognegativity}.

\begin{figure}
\centering
\includegraphics[scale=1]{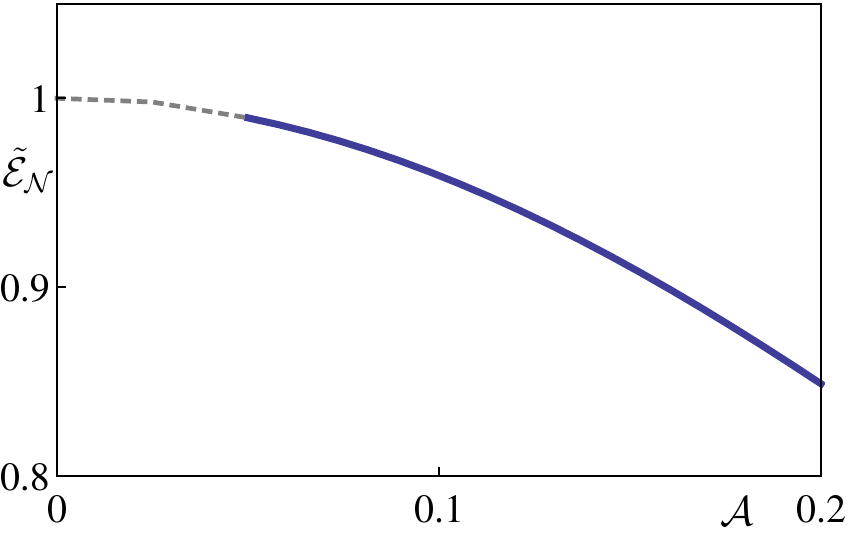}
\caption{Negativity $\tilde{\mathcal{E}}_\mathcal{N}$ of a Bell state as a function of the proper acceleration $\mathcal{A}=\mathcal{A}_\text{I}=\mathcal{A}_\text{II}$. The parameters are $m=0.1$, $\Omega_0=4.71$ and $L=2$. We normalized $\tilde{\mathcal{E}}_\mathcal{N}$ such that for vanishing accelerations $\tilde{\mathcal{E}}_\mathcal{N}=1$. The blue continuous line gives the numerical results, while the gray dotted line is an extrapolation to $\mathcal{A}=0$. From an inertial perspective Bell states are maximally entangled. However, for an accelerated observer, entanglement gets degraded with increasing acceleration.}
\label{picbell}
\end{figure}

We show the results for the entanglement of state $|B\rangle$ in Fig.\ \ref{picbell}. For vanishing acceleration the modes are maximally entangled, while the entanglement decreases with increasing acceleration. The constraint $\mathcal{A}L\lesssim 1$ that ensures that a single acceleration can be associated to a localized mode prevents us from studying arbitrarily large accelerations. 

The physical reason for the degradation of entanglement is that the overlap between the inertial wavepackets $\phi^+_k$ and the accelerated ones $ \psi_\Omega^+$ decreases with increasing acceleration. This is an inevitable effect of the Bogolyubov transformation connecting Minkowski and Rindler solutions. Therefore, with increasing acceleration the mismatch between the modes increases and the entanglement in the initially maximally entangled Bell state decreases. Interestingly, the effect of Unruh radiation is negligibly small. That is, the production of thermal particles due to the Unruh effect has only a very small effect to entanglement degradation of states of localized modes.

\section{Conclusions}\label{conclusions}

In this work, we developed a general framework to describe the effect of acceleration on arbitrary fermionic Gaussian states of two localized modes. We formulated the transformation between the quantum state seen by initial observers and the state seen by uniformly accelerated ones as the action of a fermionic Gaussian channel, and completely characterized this channel. This enabled us to study how the entanglement of the vacuum and the entanglement in Bell states is affected by acceleration.

We found that vacuum entanglement is enhanced by acceleration. That is, the correlations in the vacuum that build up due to the Unruh effect lead to entanglement between particles and antiparticles. In particular, as our framework is employing localized modes, entanglement has an operational meaning and can be extracted by suitable local detectors; a process that is sometimes referred to as ``entanglement harvesting''.  We also compare these findings  with the case of bosons \cite{Ahmadi:2016fbd} and find a qualitative agreement.

Furthermore, we quantified the entanglement degradation in a localized maximally entangled state of fermions. The degradation is due to an increasing mismatch between the initial modes and the accelerated modes that causes their overlap to decrease and, therefore, entanglement decreases as well. The effect of particle creation on entanglement degradation is negligible.

We emphasize that due to the localization of the modes, the framework presented in this work can be applied to quantum information protocols, in scenarios where local gravitational effects or effects due to acceleration are not negligible. In particular, the entanglement can be exploited as a resource.

For the future, we are interested in extending our studies to localized modes of fermions with arbitrary mutual separation, as well as studying localized Gaussian states in higher dimensional spacetimes.

\begin{acknowledgements}

B.R. thanks the Institute of Theoretical Physics at the University of Warsaw for hospitality during his visit. B.R. and Y.O. acknowledge support from Funda\c{c}\~{a}o para a Ci\^{e}ncia e a Tecnologia (Portugal), namely through programmes PTDC/POPH/POCH and projects UID/EEA/50008/2013, IT/QuSim, IT/QuNet, ProQuNet, partially funded by EU FEDER, from the EU FP7 project PAPETS (GA 323901), and from the JTF project NQN (ID 60478). Furthermore, B.R. acknowledges the support from the DP-PMI and FCT through scholarship SFRH/BD/52651/2014. K.L. and A.D. would like to acknowledge financial support from the National Science Center, Sonata BIS Grant No. DEC-2012/07/E/ST2/01402.
 
\end{acknowledgements}

\appendix

\section{Vacuum expectation values}\label{appenixcommu}

We calculate the vacuum expectation values of the Rindler creation and annihilation operators in the Minkowski vacuum. Using the transformations between the Minkowski and the Rindler vacuum given in (\ref{squeezing}), we obtain for $\langle \hat{b}_{\text{I}\Omega} \hat{b}_{\text{I}\Xi}^{\dagger}\rangle_\text{M}=\langle S^\dagger \hat{b}_{\text{I}\Omega} S S^\dagger \hat{b}_{\text{I}\Xi}^{\dagger} S\rangle_\text{R}$
\begin{equation}
\langle \hat{b}_{\text{I}\Omega} \hat{b}_{\text{I}\Xi}^{\dagger}\rangle_\text{M}= \cos^2(r_\Omega) \, \delta(\Omega-\Xi).
\end{equation}
Similarly, we obtain for $\langle \hat{b}_\Omega^{I\dagger} \hat{b}_\Theta^{I}\rangle_\text{M}=\langle S^\dagger \hat{b}_\Omega^{I\dagger} S S^\dagger \hat{b}_\Theta^{I} S\rangle_\text{R}$
\begin{equation}
\langle \hat{b}_{\text{I}\Omega}^{\dagger} \hat{b}_{\text{I}\Xi}\rangle_\text{M}=\sin^2(r_\Omega) \, \delta(\Omega-\Xi).
\end{equation}
Considering operators in different wedges, we find
\begin{equation}
\langle \hat{b}_{\text{I}\Omega} \hat{b}_{\text{II}\Xi}\rangle_\text{M}=0.
\end{equation}
Also the remaining expectation values containing only particle creation and annihilation operators are vanishing. To summarize
\begin{subequations}
\begin{align}
\langle \hat{b}_{\text{I}\Omega} \hat{b}_{\text{II}\Xi}\rangle_\text{M}=&-\langle \hat{b}_{\text{II}\Omega} \hat{b}_{\text{I}\Xi}\rangle_\text{M}=0,\\
\langle \hat{b}_{\text{I}\Omega} \hat{b}_{\text{I}\Xi}\rangle_\text{M}=&\langle \hat{b}_{\text{II}\Omega} \hat{b}_{\text{II}\Xi}\rangle_\text{M}=\langle \hat{b}_{\text{I}\Omega}^{\dagger} \hat{b}^{\dagger}_{\text{I}\Xi}\rangle_\text{M}=0,\\
\langle \hat{b}_{\text{I}\Omega}^{\dagger} \hat{b}_{\text{II}\Xi}\rangle_\text{M}=&\langle \hat{b}_{\text{II}\Omega}^{\dagger} \hat{b}_{\text{I}\Xi}\rangle_\text{M}=\langle \hat{b}_{\text{I}\Omega} \hat{b}^{\dagger}_{\text{II}\Xi}\rangle_\text{M}=0,\\
\langle \hat{b}_{\text{I}\Omega}^{\dagger} \hat{b}_{\text{I}\Xi}\rangle_\text{M}=& \langle \hat{b}_{\text{II}\Omega}^{\dagger} \hat{b}_{\text{II}\Xi}\rangle_\text{M}=\sin^2(r_\Omega)\, \delta(\Omega-\Xi),\\
\langle \hat{b}_{\text{I}\Omega} \hat{b}_{\text{I}\Xi}^{\dagger}\rangle_\text{M}=& \langle \hat{b}_{\text{II}\Omega} \hat{b}_{\text{II}\Xi}^{\dagger}\rangle_\text{M}=\cos^2(r_\Omega)\, \delta(\Omega-\Xi).
\end{align}
\end{subequations}
The same relations also hold after replacing particle by antiparticle operators. Finally, there are nonvanishing correlations between particles/antiparticles in wedge I and antiparticles/particles in wedge II. This manifests in the vacuum expectation values
\begin{align}
\langle \hat{a}_{\text{I}\Omega} \hat{b}_{\text{II}\Xi}\rangle_\text{M}=& -\langle \hat{a}_{\text{II}\Omega} \hat{b}_{\text{I}\Xi}\rangle_\text{M}=\cos(r_\Omega) \sin(r_\Omega) \delta(\Omega-\Xi),\\
\langle \hat{b}_{\text{I}\Omega} \hat{a}_{\text{II}\Xi}\rangle_\text{M}=& -\langle \hat{b}_{\text{II}\Omega} \hat{a}_{\text{I}\Xi}\rangle_\text{M}=\cos(r_\Omega) \sin(r_\Omega) \delta(\Omega-\Xi).
\end{align}

\section{Calculations of the covariance matrix}

In this appendix, we give the details of the calculations of the covariance matrices we used in this work.

\subsection{Calculation of the covariance matrix for the Minkowski vacuum}\label{appminkvac}

The covariance matrix and the first moments of the Minkowski vacuum are obtained according to (\ref{covmatrix}) and (\ref{firstmoments}), respectively. Due to the antisymmetry of the covariance matrix not all matrix elements are independent and we are left with, in general, $\frac{1}{2}n(n-1)$ independent entries for $n$ being the dimension of the matrix. A short calculation shows that, in the case of the Minkowski vacuum, the only nontrivial elements of the covariance matrix originate in terms of the kind
\begin{align}
\sigma^{(f)}_{12}=&  \langle \hat{f}_\text{I} \hat{f}_\text{I}^\dagger\rangle-\langle \hat{f}_\text{I}^\dagger \hat{f}_\text{I}\rangle.
\end{align}
The  covariance matrix of the Minkowski vacuum  is obtained to be
\begin{equation}
\sigma^{(f)}_\text{M}=\left(\begin{array}{cccc}
i\sigma_2 & 0& \dots & 0     \\
0 & i\sigma_2& \dots & 0     \\
\vdots & \vdots& \ddots & \vdots    \\
0 & 0 & 0 & i\sigma_2 \\
\end{array}\right),
\end{equation}
where $i\sigma_2=\left(\begin{array}{cc}
0 & 1   \\
-1 & 0   \\
\end{array}\right)$ and the first moments vanish, as the vacuum is an even state.

\subsection{Calculation of the covariance matrix for even Bell states}\label{appminkbell}

The covariance matrix of the even Bell state $|B\rangle$, given in (\ref{bellstate}), is derived as
\begin{align}
\sigma^{(f)}_{kl}=& 2i \langle B| \hat{c}_k \hat{c}_l |B\rangle,\hspace{10mm}\nonumber\\
=& i \langle 0| \hat{c}_k \hat{c}_l |0\rangle+ i \langle 0|\hat{f}_\text{II} \hat{f}_\text{I} \hat{c}_k \hat{c}_l \hat{f}_\text{I}^\dagger \hat{f}_\text{II}^\dagger|0\rangle+\nonumber\\
+& i \langle 0| \hat{c}_k \hat{c}_l \hat{f}_\text{I}^\dagger \hat{f}_\text{II}^\dagger|0\rangle+ i \langle 0|\hat{f}_\text{II} \hat{f}_\text{I} \hat{c}_k \hat{c}_l |0\rangle.
\end{align}
We give the calculation term by term. The first one just gives us $\frac{1}{2}$ times the vacuum matrix. Further, we need to calculate the following vacuum expectation values
\begin{align}
i\langle 0| \hat{c}_k \hat{c}_l \hat{f}_\text{I}^\dagger \hat{f}_\text{II}^\dagger|0\rangle=& \frac{i}{2}, &\text{for $k=3$, $l=1$} \nonumber\\
i\langle 0| \hat{c}_k \hat{c}_l \hat{f}_\text{I}^\dagger \hat{f}_\text{II}^\dagger|0\rangle=& -\frac{1}{2}, &\text{for $k=3$, $l=2$}\nonumber\\
i\langle 0| \hat{c}_k \hat{c}_l \hat{f}_\text{I}^\dagger \hat{f}_\text{II}^\dagger|0\rangle=& -\frac{1}{2}, &\text{for $k=4$, $l=1$}\nonumber\\
i\langle 0| \hat{c}_k \hat{c}_l \hat{f}_\text{I}^\dagger \hat{f}_\text{II}^\dagger|0\rangle=&-\frac{i}{2} , &\text{for $k=4$, $l=2$}
\end{align}
Similarly, we find
\begin{align}
i\langle 0|\hat{f}_\text{II} \hat{f}_\text{I} \hat{c}_k \hat{c}_l |0\rangle=&-\frac{i}{2} , &\text{for $k=3$, $l=1$} \nonumber\\
i\langle 0|\hat{f}_\text{II} \hat{f}_\text{I} \hat{c}_k \hat{c}_l |0\rangle=& -\frac{1}{2}, &\text{for $k=3$, $l=2$}\nonumber\\
i\langle 0|\hat{f}_\text{II} \hat{f}_\text{I} \hat{c}_k \hat{c}_l |0\rangle=& -\frac{1}{2}, &\text{for $k=4$, $l=1$}\nonumber\\
i\langle 0|\hat{f}_\text{II} \hat{f}_\text{I} \hat{c}_k \hat{c}_l |0\rangle=& \frac{i}{2}, &\text{for $k=4$, $l=2$}
\end{align} 
and
\begin{align}
i\langle 0|\hat{f}_\text{II} \hat{f}_\text{I} \hat{c}_2 \hat{c}_1 \hat{f}_\text{I}^\dagger \hat{f}_\text{II}^\dagger|0\rangle=&\frac{1}{2},
\end{align}
where the same holds for $i\langle 0|\hat{f}_\text{II} \hat{f}_\text{I} \hat{c}_4 \hat{c}_3 \hat{f}_\text{I}^\dagger \hat{f}_\text{II}^\dagger|0\rangle$ and the remaining ones are vanishing. Combining the above, we obtain for the covariance matrix of particles and their corresponding antiparticles
\begin{equation}
\sigma^{(f)}_{\text{Bell}}=\left(\begin{array}{cccccccc}
0 & 0&0 & 1  & 0 & 0& 0 & 0    \\
0 & 0&  1 & 0  & 0 & 0& 0 & 0  \\
0 & - 1& 0 & 0  & 0 & 0& 0 & 0  \\
- 1 & 0 & 0 & 0& 0 & 0& 0 & 0 \\
0 & 0 & 0 & 0 & 0 & 1& 0 & 0 \\
0 & 0 & 0 & 0 & -1 & 0& 0 & 0 \\
0 & 0 & 0 & 0 & 0 & 0& 0 &1 \\
0 & 0 & 0 & 0 & 0 & 0& -1 & 0 \\
\end{array}\right).
\end{equation}

\subsection{Calculation of the covariance matrix for the transformed Minkowski vacuum  }\label{appminkvactrafo}

In this subsection, we calculate the transformed covariance matrix $\sigma^{(d)}_{kl}= 2i \langle \hat{c}_k \hat{c}_l \rangle$. Using results for the vacuum expectations values from Appendix \ref{appenixcommu}, we find
\begin{align}\label{diagelements}
\langle \hat{d}_\text{I} \hat{d}_\text{I}^\dagger\rangle=& \int \text{d}\Omega |(\psi_\text{I}^+ , w^+_{\text{I}\Omega})|^2 \cos^2(r_\Omega).
\end{align}
Expressing (\ref{covmatrix}) in terms of the operators $\hat{d}_k$, we obtain
\begin{equation}
\sigma^{(d)}_{12}= 1-2 \int \text{d}\Omega |(\psi_\text{I}^+ , w^+_{\text{I}\Omega})|^2 \sin^2(r_\Omega).
\end{equation}
Similarly, we obtain the remaining matrix elements to be
\begin{align}
\sigma^{(d)}_{71}=& 2 \int \text{d}\Omega \,\, \Im[(\psi_\text{I}^-, w^-_{\text{I}\Omega}) (\psi_\text{II}^+, w^+_{\text{II}\Omega})]F(r_\Omega),\nonumber\\
\sigma^{(d)}_{72}=& -2 \int \text{d}\Omega \,\, \Re[(\psi_\text{I}^-, w^-_{\text{I}\Omega}) (\psi_\text{II}^+, w^+_{\text{II} \Omega})]F(r_\Omega),\nonumber\\
\sigma^{(d)}_{81}=& -2  \int \text{d}\Omega \,\, \Re[(\psi_\text{I}^-, w^-_{\text{I}\Omega}) (\psi_\text{II}^+, w^+_{\text{II} \Omega})]F(r_\Omega),\nonumber\\
\sigma^{(d)}_{82}=& -2  \int \text{d}\Omega \,\, \Im[(\psi_\text{I}^-, w^-_{\text{I}\Omega}) (\psi_\text{II}^+, w^+_{\text{II} \Omega})]F(r_\Omega),
\end{align}
where we used the definition $F(r_\Omega)=\cos(r_\Omega) \sin(r_\Omega)$. Replacing particles by antiparticles and vice versa in the above equations, one obtains the elements $\sigma^{(d)}_{53}$, $\sigma^{(d)}_{54}$, $\sigma^{(d)}_{63}$, and $\sigma^{(d)}_{64}$. Due to the antisymmetry of the covariance matrix and the symmetry between wedges I and II, these are all independent nonzero entries. Therefore, we arrive at (\ref{covmatrixnoninertial}).

\section{Logarithmic negativity}\label{lognegativity}

Given a fermionic Gaussian state $\rho$ with covariance matrix $\sigma$, we define $\Gamma=\left(\begin{array}{cc}
\Gamma_{11} & \Gamma_{12}      \\
 \Gamma_{21} & \Gamma_{22}    
\end{array}\right)=i \sigma$.
Then the partially transposed state can be written as sum of two Gaussian states $O_\pm$
\begin{equation}\label{formofptdm}
\rho^\text{pT}=\frac{1-i}{2} O_+ +\frac{1+i}{2} O_-,
\end{equation}
where the covariance matrices of the $O_\pm$ are $\Gamma_+=\left(\begin{array}{cc}
\Gamma_{11} & i\Gamma_{12}      \\
 i\Gamma_{21} & -\Gamma_{22}    
\end{array}\right)$
and $\Gamma_-=\left(\begin{array}{cc}
\Gamma_{11} &- i\Gamma_{12}      \\
 -i\Gamma_{21} & -\Gamma_{22}    
\end{array}\right)$,
respectively \cite{eisler2015partial}. If $[\Gamma_+, \Gamma_-]=0$ holds, the logarithmic negativity $\mathcal{E}$ can be calculated exactly. One example of such states are the isotropic states for which $\Gamma^2=-\lambda^2 1$ holds \cite{Alonso2004}. As we see that does not hold for the transformed vacuum state.

We consider the vacuum covariance matrix (\ref{covmatrixnoninertialdntriv}) that we rescale by $i$, i.e., $\Gamma^{(\text{vac})}=i \sigma^{(d)}_{}$. To obtain the logarithmic negativity we have to partially transpose the density matrix. It is of the form (\ref{formofptdm}). Accordingly, we define $\Gamma_+$ and $\Gamma_-$ as above. Then calculating the commutator $[\Gamma_+, \Gamma_-]$, we find $[\Gamma_+, \Gamma_-]\neq 0$ and we cannot calculate the logarithmic negativity exactly. If $O_+$ and $O_-$ commute they can be diagonalized simultaneously and we find that the eigenvalues of $O_+$ and $O_-$ are complex conjugate to each other. What is left to do is to find the eigenvalues of $O_+$ given $\Gamma_+$. We denote the eigenvalues of $\Gamma_+$ by $\pm\nu^s$ and write $O_+$ as
\begin{align}
O_+=&\prod_{s=\pm}\frac{1+i\nu^s \hat{h}_{1}^s \hat{h}_{2}^s}{2},
\end{align}
where the $\hat{h}_j^\pm$ are the Majorana operators obtained from the $\hat{c}_j$ via the operation diagonalizing $\Gamma_+$. Then the eigenvalues of $O_+$ are given by
\begin{equation}
\omega^{s s'}=\frac{1}{4}(1+s \nu^+)(1+s' \nu^-).
\end{equation}
To study the vacuum entanglement, we have first to find the eigenvalues $\pm\nu^s$ of $\Gamma_+^{}$, where we restrict ourselves to the entanglement between particles I and antiparticles II (the entanglement between antiparticles I and particles II is analogous). $\Gamma_+^{(\text{vac})}$ reads
\begin{equation}\label{gammavacd0}
\left(\begin{array}{cccc}
0 & iN_\text{I}^+& -\Im[N_\text{I,II}^+] & -\Re[N_\text{I,II}^+]      \\
-iN_\text{I}^+ & 0&  -\Re[N_\text{I,II}^+] & \Im[N_\text{I,II}^+]     \\
\Im[N_\text{I,II}^+] & \Re[N_\text{I,II}^+] & 0 & -iN_\text{II}^-      \\
\Re[N_\text{I,II}^+] & -\Im[N_\text{I,II}^+] & iN_\text{II}^- & 0
\end{array}\right).
\end{equation}
The eigenvalues of (\ref{gammavacd0}) are given by
\begin{subequations}\label{vaceigenvaluesd0}
\begin{align}
\pm\nu^+=&\pm\frac{1}{2}\left(N_\text{I}^++N_\text{II}^-+\sqrt{(N_\text{I}^+-N_\text{II}^-)^2-4|N_\text{I,II}^+|^2}\right),\\
\pm\nu^-=&\pm\frac{1}{2}\left(N_\text{I}^++N_\text{II}^--\sqrt{(N_\text{I}^+-N_\text{II}^-)^2-4|N_\text{I,II}^+|^2}\right).
\end{align}
\end{subequations}
From these we can write the eigenvalues of $O_+$ as
\begin{align}
\omega_{++}=&\frac{1}{4}(1+\nu^++\nu^-+\nu^+\nu^-), \nonumber\\
\omega_{--}=&\frac{1}{4}(1-\nu^+-\nu^-+\nu^+\nu^-),\nonumber\\
\omega_{+-}=&\frac{1}{4}(1+\nu^+-\nu^--\nu^+\nu^-), \nonumber\\
\omega_{-+}=&\frac{1}{4}(1-\nu^++\nu^--\nu^+\nu^-)
\end{align}
and a lower bound for the entanglement  can  be obtained \cite{eisler2015partial}. It is given by
\begin{align}\label{lowerboundneg}
\mathcal{E}_\mathcal{N}\geq\tilde{\mathcal{E}}_\mathcal{N}=&\ln\left(1-2\,\text{Tr}_o \,\rho^\text{pT} \right)\nonumber\\
=&\ln\left(\,\text{Tr}_e \,\rho^\text{pT}-\text{Tr}_o \,\rho^\text{pT}\right)\nonumber\\
=&\ln(\Re[\,\text{Tr}_e \,O_+-\text{Tr}_o \,O_+]+\nonumber\\
+& \Im[\,\text{Tr}_e \,O_+-\text{Tr}_o \,O_+]),
\end{align}
where $\text{Tr}_{e/o}$ denotes the trace over the even and odd subspaces, respectively. We next calculate $\text{Tr}_e \,O_+-\text{Tr}_o \,O_+$ that can be given in terms of the eigenvalues $\omega$ as
\begin{equation}
\text{Tr}_e \,O_+-\text{Tr}_o \,O_+=\sum_{s,s'} S_{s,s'}\omega^{s s'}
\end{equation}
with
\begin{equation}
S_{s,s'}=\Re[l_{s,s'}]+\Im[l_{s,s'}],
\end{equation}
where $l_{s,s'}=1$ for $s=s'$ and $l_{s,s'}=is$ for $s=-s'$. Thus, we find
\begin{align}
\text{Tr}_e \,O_+-\text{Tr}_o \,O_+=&\sum_{s,s'} S_{s,s'}\omega^{s s'}\nonumber\\
=& \frac{1}{2}(1+\nu^+\nu^-+\nu^+-\nu^-)
\end{align}
and, therefore, we obtain the lower bound
\begin{align}
\tilde{\mathcal{E}}_\mathcal{N}=&\ln(\frac{1}{2}(1+N_\text{I}^+N_\text{II}^-+|N_\text{I,II}^+|^2  +\nonumber\\
+& \Re[\sqrt{(N_\text{I}^+-N_\text{II}^-)^2-4|N_\text{I,II}^+|^2}]+\nonumber\\
+& \Im[\sqrt{(N_\text{I}^+-N_\text{II}^-)^2-4|N_\text{I,II}^+|^2}] ))
\end{align}
for the entanglement between two  modes in the vacuum.


\begin{thebibliography}{31}%
\makeatletter
\providecommand \@ifxundefined [1]{%
 \@ifx{#1\undefined}
}%
\providecommand \@ifnum [1]{%
 \ifnum #1\expandafter \@firstoftwo
 \else \expandafter \@secondoftwo
 \fi
}%
\providecommand \@ifx [1]{%
 \ifx #1\expandafter \@firstoftwo
 \else \expandafter \@secondoftwo
 \fi
}%
\providecommand \natexlab [1]{#1}%
\providecommand \enquote  [1]{``#1''}%
\providecommand \bibnamefont  [1]{#1}%
\providecommand \bibfnamefont [1]{#1}%
\providecommand \citenamefont [1]{#1}%
\providecommand \href@noop [0]{\@secondoftwo}%
\providecommand \href [0]{\begingroup \@sanitize@url \@href}%
\providecommand \@href[1]{\@@startlink{#1}\@@href}%
\providecommand \@@href[1]{\endgroup#1\@@endlink}%
\providecommand \@sanitize@url [0]{\catcode `\\12\catcode `\$12\catcode
  `\&12\catcode `\#12\catcode `\^12\catcode `\_12\catcode `\%12\relax}%
\providecommand \@@startlink[1]{}%
\providecommand \@@endlink[0]{}%
\providecommand \url  [0]{\begingroup\@sanitize@url \@url }%
\providecommand \@url [1]{\endgroup\@href {#1}{\urlprefix }}%
\providecommand \urlprefix  [0]{URL }%
\providecommand \Eprint [0]{\href }%
\providecommand \doibase [0]{http://dx.doi.org/}%
\providecommand \selectlanguage [0]{\@gobble}%
\providecommand \bibinfo  [0]{\@secondoftwo}%
\providecommand \bibfield  [0]{\@secondoftwo}%
\providecommand \translation [1]{[#1]}%
\providecommand \BibitemOpen [0]{}%
\providecommand \bibitemStop [0]{}%
\providecommand \bibitemNoStop [0]{.\EOS\space}%
\providecommand \EOS [0]{\spacefactor3000\relax}%
\providecommand \BibitemShut  [1]{\csname bibitem#1\endcsname}%
\let\auto@bib@innerbib\@empty
\bibitem [{\citenamefont {Unruh}(1976)}]{Unruh:1976db}%
  \BibitemOpen
  \bibfield  {author} {\bibinfo {author} {\bibfnamefont {W.~G.}\ \bibnamefont
  {Unruh}},\ }\href {\doibase 10.1103/PhysRevD.14.870} {\bibfield  {journal}
  {\bibinfo  {journal} {Phys. Rev.}\ }\textbf {\bibinfo {volume} {D 14}},\
  \bibinfo {pages} {870} (\bibinfo {year} {1976})}\BibitemShut {NoStop}%
\bibitem [{\citenamefont {Peres}\ and\ \citenamefont
  {Terno}(2004)}]{Peres:2002wx}%
  \BibitemOpen
  \bibfield  {author} {\bibinfo {author} {\bibfnamefont {A.}~\bibnamefont
  {Peres}}\ and\ \bibinfo {author} {\bibfnamefont {D.~R.}\ \bibnamefont
  {Terno}},\ }\href {\doibase 10.1103/RevModPhys.76.93} {\bibfield  {journal}
  {\bibinfo  {journal} {Rev. Mod. Phys.}\ }\textbf {\bibinfo {volume} {76}},\
  \bibinfo {pages} {93} (\bibinfo {year} {2004})},\ \Eprint
  {http://arxiv.org/abs/quant-ph/0212023} {arXiv:quant-ph/0212023 [quant-ph]}
  \BibitemShut {NoStop}%
\bibitem [{\citenamefont {Fuentes-Schuller}\ and\ \citenamefont
  {Mann}(2005)}]{FuentesSchuller:2004xp}%
  \BibitemOpen
  \bibfield  {author} {\bibinfo {author} {\bibfnamefont {I.}~\bibnamefont
  {Fuentes-Schuller}}\ and\ \bibinfo {author} {\bibfnamefont {R.~B.}\
  \bibnamefont {Mann}},\ }\href {\doibase 10.1103/PhysRevLett.95.120404}
  {\bibfield  {journal} {\bibinfo  {journal} {Phys. Rev. Lett.}\ }\textbf
  {\bibinfo {volume} {95}},\ \bibinfo {pages} {120404} (\bibinfo {year}
  {2005})},\ \Eprint {http://arxiv.org/abs/quant-ph/0410172}
  {arXiv:quant-ph/0410172 [quant-ph]} \BibitemShut {NoStop}%
\bibitem [{\citenamefont {Bruschi}\ \emph {et~al.}(2010)\citenamefont
  {Bruschi}, \citenamefont {Louko}, \citenamefont {Martin-Martinez},
  \citenamefont {Dragan},\ and\ \citenamefont {Fuentes}}]{Bruschi:2010mc}%
  \BibitemOpen
  \bibfield  {author} {\bibinfo {author} {\bibfnamefont {D.~E.}\ \bibnamefont
  {Bruschi}}, \bibinfo {author} {\bibfnamefont {J.}~\bibnamefont {Louko}},
  \bibinfo {author} {\bibfnamefont {E.}~\bibnamefont {Martin-Martinez}},
  \bibinfo {author} {\bibfnamefont {A.}~\bibnamefont {Dragan}}, \ and\ \bibinfo
  {author} {\bibfnamefont {I.}~\bibnamefont {Fuentes}},\ }\href {\doibase
  10.1103/PhysRevA.82.042332} {\bibfield  {journal} {\bibinfo  {journal} {Phys.
  Rev.}\ }\textbf {\bibinfo {volume} {A 82}},\ \bibinfo {pages} {042332}
  (\bibinfo {year} {2010})},\ \Eprint {http://arxiv.org/abs/1007.4670}
  {arXiv:1007.4670 [quant-ph]} \BibitemShut {NoStop}%
\bibitem [{\citenamefont {Richter}\ and\ \citenamefont
  {Omar}(2015)}]{Richter:2015wha}%
  \BibitemOpen
  \bibfield  {author} {\bibinfo {author} {\bibfnamefont {B.}~\bibnamefont
  {Richter}}\ and\ \bibinfo {author} {\bibfnamefont {Y.}~\bibnamefont {Omar}},\
  }\href {\doibase 10.1103/PhysRevA.92.022334} {\bibfield  {journal} {\bibinfo
  {journal} {Phys. Rev.}\ }\textbf {\bibinfo {volume} {A 92}},\ \bibinfo
  {pages} {022334} (\bibinfo {year} {2015})},\ \Eprint
  {http://arxiv.org/abs/1503.07526} {arXiv:1503.07526 [quant-ph]} \BibitemShut
  {NoStop}%
\bibitem [{\citenamefont {Dragan}\ \emph
  {et~al.}(2013{\natexlab{a}})\citenamefont {Dragan}, \citenamefont {Doukas},
  \citenamefont {Martin-Martinez},\ and\ \citenamefont
  {Bruschi}}]{Dragan:2012hy}%
  \BibitemOpen
  \bibfield  {author} {\bibinfo {author} {\bibfnamefont {A.}~\bibnamefont
  {Dragan}}, \bibinfo {author} {\bibfnamefont {J.}~\bibnamefont {Doukas}},
  \bibinfo {author} {\bibfnamefont {E.}~\bibnamefont {Martin-Martinez}}, \ and\
  \bibinfo {author} {\bibfnamefont {D.~E.}\ \bibnamefont {Bruschi}},\ }\href
  {\doibase 10.1088/0264-9381/30/23/235006} {\bibfield  {journal} {\bibinfo
  {journal} {Class. Quant. Grav.}\ }\textbf {\bibinfo {volume} {30}},\ \bibinfo
  {pages} {235006} (\bibinfo {year} {2013}{\natexlab{a}})},\ \Eprint
  {http://arxiv.org/abs/1203.0655} {arXiv:1203.0655 [quant-ph]} \BibitemShut
  {NoStop}%
\bibitem [{\citenamefont {Ahmadi}\ \emph {et~al.}(2016)\citenamefont {Ahmadi},
  \citenamefont {Lorek}, \citenamefont {Ch\c{e}ci\'{n}ska}, \citenamefont
  {Smith}, \citenamefont {Mann},\ and\ \citenamefont
  {Dragan}}]{Ahmadi:2016fbd}%
  \BibitemOpen
  \bibfield  {author} {\bibinfo {author} {\bibfnamefont {M.}~\bibnamefont
  {Ahmadi}}, \bibinfo {author} {\bibfnamefont {K.}~\bibnamefont {Lorek}},
  \bibinfo {author} {\bibfnamefont {A.}~\bibnamefont {Ch\c{e}ci\'{n}ska}},
  \bibinfo {author} {\bibfnamefont {A.~R.~H.}\ \bibnamefont {Smith}}, \bibinfo
  {author} {\bibfnamefont {R.~B.}\ \bibnamefont {Mann}}, \ and\ \bibinfo
  {author} {\bibfnamefont {A.}~\bibnamefont {Dragan}},\ }\href {\doibase
  10.1103/PhysRevD.93.124031} {\bibfield  {journal} {\bibinfo  {journal} {Phys.
  Rev.}\ }\textbf {\bibinfo {volume} {D 93}},\ \bibinfo {pages} {124031}
  (\bibinfo {year} {2016})},\ \Eprint {http://arxiv.org/abs/1602.02349}
  {arXiv:1602.02349 [quant-ph]} \BibitemShut {NoStop}%
\bibitem [{\citenamefont {Birrell}\ and\ \citenamefont
  {Davies}(1984)}]{birrell1984quantum}%
  \BibitemOpen
  \bibfield  {author} {\bibinfo {author} {\bibfnamefont {N.~D.}\ \bibnamefont
  {Birrell}}\ and\ \bibinfo {author} {\bibfnamefont {P.~C.~W.}\ \bibnamefont
  {Davies}},\ }\href@noop {} {\emph {\bibinfo {title} {{Quantum fields in
  curved space}}}}\ (\bibinfo  {publisher} {Cambridge university press},\
  \bibinfo {year} {1984})\BibitemShut {NoStop}%
\bibitem [{\citenamefont {Benatti}\ and\ \citenamefont
  {Floreanini}(2004)}]{benatti2004entanglement}%
  \BibitemOpen
  \bibfield  {author} {\bibinfo {author} {\bibfnamefont {F.}~\bibnamefont
  {Benatti}}\ and\ \bibinfo {author} {\bibfnamefont {R.}~\bibnamefont
  {Floreanini}},\ }\href {\doibase 10.1103/PhysRevA.70.012112} {\bibfield  {journal} {\bibinfo  {journal}
  {Phys. Rev.}\ }\textbf {\bibinfo {volume} {A 70}},\ \bibinfo {pages} {012112}
  (\bibinfo {year} {2004})},\ \Eprint {http://arxiv.org/abs/quant-ph/0403157}
  {arXiv:quant-ph/0403157 [quant-ph]} \BibitemShut {NoStop}%
 \bibitem [{\citenamefont {Salton}\ \emph {et~al.}(2015)\citenamefont {Salton},
  \citenamefont {Mann},\ and\ \citenamefont {Menicucci}}]{Salton:2014jaa}%
  \BibitemOpen
  \bibfield  {author} {\bibinfo {author} {\bibfnamefont {G.}~\bibnamefont
  {Salton}}, \bibinfo {author} {\bibfnamefont {R.~B.}\ \bibnamefont {Mann}}, \
  and\ \bibinfo {author} {\bibfnamefont {N.~C.}\ \bibnamefont {Menicucci}},\
  }\href {\doibase 10.1088/1367-2630/17/3/035001} {\bibfield  {journal}
  {\bibinfo  {journal} {New J. Phys.}\ }\textbf {\bibinfo {volume} {17}},\
  \bibinfo {pages} {035001} (\bibinfo {year} {2015})},\ \Eprint
  {http://arxiv.org/abs/1408.1395} {arXiv:1408.1395 [quant-ph]} \BibitemShut
  {NoStop}%
\bibitem [{\citenamefont {Lorek}\ \emph {et~al.}(2014)\citenamefont {Lorek},
  \citenamefont {Pecak}, \citenamefont {Brown},\ and\ \citenamefont
  {Dragan}}]{Lorek:2014dwa}%
  \BibitemOpen
  \bibfield  {author} {\bibinfo {author} {\bibfnamefont {K.}~\bibnamefont
  {Lorek}}, \bibinfo {author} {\bibfnamefont {D.}~\bibnamefont {Pecak}},
  \bibinfo {author} {\bibfnamefont {E.~G.}\ \bibnamefont {Brown}}, \ and\
  \bibinfo {author} {\bibfnamefont {A.}~\bibnamefont {Dragan}},\ }\href
  {\doibase 10.1103/PhysRevA.90.032316} {\bibfield  {journal} {\bibinfo
  {journal} {Phys. Rev.}\ }\textbf {\bibinfo {volume} {A 90}},\ \bibinfo
  {pages} {032316} (\bibinfo {year} {2014})}\BibitemShut {NoStop}%
\bibitem [{\citenamefont {Martin-Martinez}\ and\ \citenamefont
  {Menicucci}(2012)}]{MartinMartinez:2012sg}%
  \BibitemOpen
  \bibfield  {author} {\bibinfo {author} {\bibfnamefont {E.}~\bibnamefont
  {Martin-Martinez}}\ and\ \bibinfo {author} {\bibfnamefont {N.~C.}\
  \bibnamefont {Menicucci}},\ }\href {\doibase 10.1088/0264-9381/29/22/224003}
  {\bibfield  {journal} {\bibinfo  {journal} {Class. Quant. Grav.}\ }\textbf
  {\bibinfo {volume} {29}},\ \bibinfo {pages} {224003} (\bibinfo {year}
  {2012})},\ \Eprint {http://arxiv.org/abs/1204.4918} {arXiv:1204.4918 [gr-qc]}
  \BibitemShut {NoStop}%
\bibitem [{\citenamefont {del Rey}\ \emph {et~al.}(2012)\citenamefont {del
  Rey}, \citenamefont {Porras},\ and\ \citenamefont
  {Martin-Martinez}}]{delRey:2011jt}%
  \BibitemOpen
  \bibfield  {author} {\bibinfo {author} {\bibfnamefont {M.}~\bibnamefont {del
  Rey}}, \bibinfo {author} {\bibfnamefont {D.}~\bibnamefont {Porras}}, \ and\
  \bibinfo {author} {\bibfnamefont {E.}~\bibnamefont {Martin-Martinez}},\
  }\href {\doibase 10.1103/PhysRevA.85.022511} {\bibfield  {journal} {\bibinfo
  {journal} {Phys. Rev.}\ }\textbf {\bibinfo {volume} {A 85}},\ \bibinfo
  {pages} {022511} (\bibinfo {year} {2012})},\ \Eprint
  {http://arxiv.org/abs/1109.0209} {arXiv:1109.0209 [quant-ph]} \BibitemShut
  {NoStop}%
  \bibitem [{\citenamefont {Garc{\'\i}a-{\'A}lvarez}\ \emph
  {et~al.}(2016)\citenamefont {Garc{\'\i}a-{\'A}lvarez}, \citenamefont
  {Felicetti}, \citenamefont {Rico}, \citenamefont {Solano},\ and\
  \citenamefont {Sab{\'\i}n}}]{garcia2016entanglement}%
  \BibitemOpen
  \bibfield  {author} {\bibinfo {author} {\bibfnamefont {L.}~\bibnamefont
  {Garc{\'\i}a-{\'A}lvarez}}, \bibinfo {author} {\bibfnamefont
  {S.}~\bibnamefont {Felicetti}}, \bibinfo {author} {\bibfnamefont
  {E.}~\bibnamefont {Rico}}, \bibinfo {author} {\bibfnamefont {E.}~\bibnamefont
  {Solano}}, \ and\ \bibinfo {author} {\bibfnamefont {C.}~\bibnamefont
  {Sab{\'\i}n}},\ }\href
  {http://dx.doi.org/10.1038/s41598-017-00770-z} {\bibfield  {journal} {\bibinfo
  {journal} {Sci. Rep.}\ }\textbf {\bibinfo {volume} {7}},\ \bibinfo
  {pages} {657} (\bibinfo {year} {2017})}\BibitemShut {NoStop}%
\bibitem [{\citenamefont {Richter}\ \emph {et~al.}(2017)\citenamefont
  {Richter}, \citenamefont {Terças}, \citenamefont {Omar},\ and\ \citenamefont
  {de~Vega}}]{richter2017}%
  \BibitemOpen
  \bibfield  {author} {\bibinfo {author} {\bibfnamefont {B.}~\bibnamefont
  {Richter}}, \bibinfo {author} {\bibfnamefont {H.}~\bibnamefont {Terças}},
  \bibinfo {author} {\bibfnamefont {Y.}~\bibnamefont {Omar}}, \ and\ \bibinfo
  {author} {\bibfnamefont {I.}~\bibnamefont {de~Vega}},\ }\href@noop {} {\
  (\bibinfo {year} {2017})},\ \Eprint {http://arxiv.org/abs/1705.00008}
  {arXiv:1705.00008 [quant-ph]} \BibitemShut {NoStop}%
\bibitem [{\citenamefont {Kubicki}\ \emph {et~al.}(2016)\citenamefont
  {Kubicki}, \citenamefont {Westman},\ and\ \citenamefont
  {Leon}}]{Kubicki:2016ngz}%
  \BibitemOpen
  \bibfield  {author} {\bibinfo {author} {\bibfnamefont {A.~M.}\ \bibnamefont
  {Kubicki}}, \bibinfo {author} {\bibfnamefont {H.}~\bibnamefont {Westman}}, \
  and\ \bibinfo {author} {\bibfnamefont {J.}~\bibnamefont {Leon}},\ }\href@noop
  {} {\  (\bibinfo {year} {2016})},\ \Eprint {http://arxiv.org/abs/1606.03286}
  {arXiv:1606.03286 [quant-ph]} \BibitemShut {NoStop}%
\bibitem [{\citenamefont {Dragan}\ \emph
  {et~al.}(2013{\natexlab{b}})\citenamefont {Dragan}, \citenamefont {Doukas},\
  and\ \citenamefont {Martín-Martínez}}]{Dragan:2013dna}%
  \BibitemOpen
  \bibfield  {author} {\bibinfo {author} {\bibfnamefont {A.}~\bibnamefont
  {Dragan}}, \bibinfo {author} {\bibfnamefont {J.}~\bibnamefont {Doukas}}, \
  and\ \bibinfo {author} {\bibfnamefont {E.}~\bibnamefont
  {Martín-Martínez}},\ }\href {\doibase 10.1103/PhysRevA.87.052326}
  {\bibfield  {journal} {\bibinfo  {journal} {Phys. Rev.}\ }\textbf {\bibinfo
  {volume} {A 87}},\ \bibinfo {pages} {052326} (\bibinfo {year}
  {2013}{\natexlab{b}})},\ \Eprint {http://arxiv.org/abs/1207.4275}
  {arXiv:1207.4275} \BibitemShut {NoStop}%
\bibitem [{\citenamefont {Bravyi}(2005{\natexlab{a}})}]{Bravyi05}%
  \BibitemOpen
  \bibfield  {author} {\bibinfo {author} {\bibfnamefont {S.}~\bibnamefont
  {Bravyi}},\ }\href {http://dl.acm.org/citation.cfm?id=2011637.2011640}
  {\bibfield  {journal} {\bibinfo  {journal} {Quantum Info. Comput.}\ }\textbf
  {\bibinfo {volume} {5}},\ \bibinfo {pages} {216} (\bibinfo {year}
  {2005}{\natexlab{a}})}\BibitemShut {NoStop}%
\bibitem [{\citenamefont {Holevo}\ and\ \citenamefont
  {Werner}(2001)}]{PhysRevA.63.032312}%
  \BibitemOpen
  \bibfield  {author} {\bibinfo {author} {\bibfnamefont {A.~S.}\ \bibnamefont
  {Holevo}}\ and\ \bibinfo {author} {\bibfnamefont {R.~F.}\ \bibnamefont
  {Werner}},\ }\href {\doibase 10.1103/PhysRevA.63.032312} {\bibfield
  {journal} {\bibinfo  {journal} {Phys. Rev.}\ }\textbf {\bibinfo {volume} {A
  63}},\ \bibinfo {pages} {032312} (\bibinfo {year} {2001})}\BibitemShut
  {NoStop}%
\bibitem [{\citenamefont {Braunstein}\ and\ \citenamefont {van
  Loock}(2005)}]{RevModPhys.77.513}%
  \BibitemOpen
  \bibfield  {author} {\bibinfo {author} {\bibfnamefont {S.~L.}\ \bibnamefont
  {Braunstein}}\ and\ \bibinfo {author} {\bibfnamefont {P.}~\bibnamefont {van
  Loock}},\ }\href {\doibase 10.1103/RevModPhys.77.513} {\bibfield  {journal}
  {\bibinfo  {journal} {Rev. Mod. Phys.}\ }\textbf {\bibinfo {volume} {77}},\
  \bibinfo {pages} {513} (\bibinfo {year} {2005})}\BibitemShut {NoStop}%
\bibitem [{\citenamefont {Weedbrook}\ \emph {et~al.}(2012)\citenamefont
  {Weedbrook}, \citenamefont {Pirandola}, \citenamefont {Garc\'{\i}a-Patr\'on},
  \citenamefont {Cerf}, \citenamefont {Ralph}, \citenamefont {Shapiro},\ and\
  \citenamefont {Lloyd}}]{RevModPhys.84.621}%
  \BibitemOpen
  \bibfield  {author} {\bibinfo {author} {\bibfnamefont {C.}~\bibnamefont
  {Weedbrook}}, \bibinfo {author} {\bibfnamefont {S.}~\bibnamefont
  {Pirandola}}, \bibinfo {author} {\bibfnamefont {R.}~\bibnamefont
  {Garc\'{\i}a-Patr\'on}}, \bibinfo {author} {\bibfnamefont {N.~J.}\
  \bibnamefont {Cerf}}, \bibinfo {author} {\bibfnamefont {T.~C.}\ \bibnamefont
  {Ralph}}, \bibinfo {author} {\bibfnamefont {J.~H.}\ \bibnamefont {Shapiro}},
  \ and\ \bibinfo {author} {\bibfnamefont {S.}~\bibnamefont {Lloyd}},\ }\href
  {\doibase 10.1103/RevModPhys.84.621} {\bibfield  {journal} {\bibinfo
  {journal} {Rev. Mod. Phys.}\ }\textbf {\bibinfo {volume} {84}},\ \bibinfo
  {pages} {621} (\bibinfo {year} {2012})}\BibitemShut {NoStop}%
\bibitem [{\citenamefont {Bloch}\ and\ \citenamefont
  {Messiah}(1962)}]{bloch1962canonical}%
  \BibitemOpen
  \bibfield  {author} {\bibinfo {author} {\bibfnamefont {C.}~\bibnamefont
  {Bloch}}\ and\ \bibinfo {author} {\bibfnamefont {A.}~\bibnamefont
  {Messiah}},\ }\href {\doibase 10.1016/0029-5582(62)90377-2} {\bibfield
  {journal} {\bibinfo  {journal} {Nucl. Phys.}\ }\textbf {\bibinfo {volume}
  {39}},\ \bibinfo {pages} {95} (\bibinfo {year} {1962})}\BibitemShut {NoStop}%
\bibitem [{\citenamefont {Kraus}\ \emph {et~al.}(2009)\citenamefont {Kraus},
  \citenamefont {Wolf}, \citenamefont {Cirac},\ and\ \citenamefont
  {Giedke}}]{kraus2009pairing}%
  \BibitemOpen
  \bibfield  {author} {\bibinfo {author} {\bibfnamefont {C.~V.}\ \bibnamefont
  {Kraus}}, \bibinfo {author} {\bibfnamefont {M.~M.}\ \bibnamefont {Wolf}},
  \bibinfo {author} {\bibfnamefont {J.~I.}\ \bibnamefont {Cirac}}, \ and\
  \bibinfo {author} {\bibfnamefont {G.}~\bibnamefont {Giedke}},\ }\href
  {\doibase 10.1103/PhysRevA.79.012306} {\bibfield  {journal} {\bibinfo
  {journal} {Phys. Rev.}\ }\textbf {\bibinfo {volume} {A 79}},\ \bibinfo
  {pages} {012306} (\bibinfo {year} {2009})}\BibitemShut {NoStop}%
\bibitem [{\citenamefont {de~Melo}\ \emph {et~al.}(2013)\citenamefont
  {de~Melo}, \citenamefont {{\'C}wikli{\'n}ski},\ and\ \citenamefont
  {Terhal}}]{de2013power}%
  \BibitemOpen
  \bibfield  {author} {\bibinfo {author} {\bibfnamefont {F.}~\bibnamefont
  {de~Melo}}, \bibinfo {author} {\bibfnamefont {P.}~\bibnamefont
  {{\'C}wikli{\'n}ski}}, \ and\ \bibinfo {author} {\bibfnamefont {B.~M.}\
  \bibnamefont {Terhal}},\ }\href
  {http://stacks.iop.org/1367-2630/15/i=1/a=013015} {\bibfield  {journal}
  {\bibinfo  {journal} {New J. Phys.}\ }\textbf {\bibinfo {volume}
  {15}},\ \bibinfo {pages} {013015} (\bibinfo {year} {2013})}\BibitemShut
  {NoStop}%
\bibitem [{\citenamefont {Bravyi}(2005{\natexlab{b}})}]{bravyi2005classical}%
  \BibitemOpen
  \bibfield  {author} {\bibinfo {author} {\bibfnamefont {S.}~\bibnamefont
  {Bravyi}},\ }\href@noop {} {\  (\bibinfo {year} {2005}{\natexlab{b}})},\
  \Eprint {http://arxiv.org/abs/quant-ph/0507282} {quant-ph/0507282}
  \BibitemShut {NoStop}%
\bibitem [{\citenamefont {Bradler}\ \emph {et~al.}(2011)\citenamefont
  {Bradler}, \citenamefont {Jochym-O'Connor},\ and\ \citenamefont
  {Jauregui}}]{Bradler:2010st}%
  \BibitemOpen
  \bibfield  {author} {\bibinfo {author} {\bibfnamefont {K.}~\bibnamefont
  {Bradler}}, \bibinfo {author} {\bibfnamefont {T.}~\bibnamefont
  {Jochym-O'Connor}}, \ and\ \bibinfo {author} {\bibfnamefont {R.}~\bibnamefont
  {Jauregui}},\ }\href {\doibase 10.1063/1.3597233} {\bibfield  {journal}
  {\bibinfo  {journal} {J. Math. Phys.}\ }\textbf {\bibinfo {volume} {52}},\
  \bibinfo {pages} {062202} (\bibinfo {year} {2011})},\ \Eprint
  {http://arxiv.org/abs/1011.2215} {arXiv:1011.2215 [quant-ph]} \BibitemShut
  {NoStop}%
\bibitem [{\citenamefont {Greplov{\'a}}\ and\ \citenamefont
  {Giedke}(2016)}]{greplova2016degradability}%
  \BibitemOpen
  \bibfield  {author} {\bibinfo {author} {\bibfnamefont {E.}~\bibnamefont
  {Greplov{\'a}}}\ and\ \bibinfo {author} {\bibfnamefont {G.}~\bibnamefont
  {Giedke}},\ }\href@noop {} {\  (\bibinfo {year} {2016})},\ \Eprint
  {http://arxiv.org/abs/1604.01954} {arXiv:1604.01954 [quant-ph]} \BibitemShut
  {NoStop}%
\bibitem [{\citenamefont {Crispino}\ \emph {et~al.}(2008)\citenamefont
  {Crispino}, \citenamefont {Higuchi},\ and\ \citenamefont
  {Matsas}}]{Crispino:2007eb}%
  \BibitemOpen
  \bibfield  {author} {\bibinfo {author} {\bibfnamefont {L.~C.~B.}\
  \bibnamefont {Crispino}}, \bibinfo {author} {\bibfnamefont {A.}~\bibnamefont
  {Higuchi}}, \ and\ \bibinfo {author} {\bibfnamefont {G.~E.~A.}\ \bibnamefont
  {Matsas}},\ }\href {\doibase 10.1103/RevModPhys.80.787} {\bibfield  {journal}
  {\bibinfo  {journal} {Rev. Mod. Phys.}\ }\textbf {\bibinfo {volume} {80}},\
  \bibinfo {pages} {787} (\bibinfo {year} {2008})},\ \Eprint
  {http://arxiv.org/abs/0710.5373} {arXiv:0710.5373 [gr-qc]} \BibitemShut
  {NoStop}%
\bibitem [{\citenamefont {Takagi}(1986)}]{takagi1986vacuum}%
  \BibitemOpen
  \bibfield  {author} {\bibinfo {author} {\bibfnamefont {S.}~\bibnamefont
  {Takagi}},\ }\href@noop {} {\bibfield  {journal} {\bibinfo  {journal} {Prog.
  Theor. Phys. Suppl.}\ }\textbf {\bibinfo {volume} {88}},\ \bibinfo {pages}
  {1} (\bibinfo {year} {1986})}\BibitemShut {NoStop}%
\bibitem [{\citenamefont {Friis}\ \emph {et~al.}(2013)\citenamefont {Friis},
  \citenamefont {Lee},\ and\ \citenamefont {Louko}}]{Friis:2013eva}%
  \BibitemOpen
  \bibfield  {author} {\bibinfo {author} {\bibfnamefont {N.}~\bibnamefont
  {Friis}}, \bibinfo {author} {\bibfnamefont {A.~R.}\ \bibnamefont {Lee}}, \
  and\ \bibinfo {author} {\bibfnamefont {J.}~\bibnamefont {Louko}},\ }\href
  {\doibase 10.1103/PhysRevD.88.064028} {\bibfield  {journal} {\bibinfo
  {journal} {Phys. Rev.}\ }\textbf {\bibinfo {volume} {D 88}},\ \bibinfo
  {pages} {064028} (\bibinfo {year} {2013})},\ \Eprint
  {http://arxiv.org/abs/1307.1631} {arXiv:1307.1631 [quant-ph]} \BibitemShut
  {NoStop}%
\bibitem [{\citenamefont {Eisler}\ and\ \citenamefont
  {Zimbor{\'a}s}(2015)}]{eisler2015partial}%
  \BibitemOpen
  \bibfield  {author} {\bibinfo {author} {\bibfnamefont {V.}~\bibnamefont
  {Eisler}}\ and\ \bibinfo {author} {\bibfnamefont {Z.}~\bibnamefont
  {Zimbor{\'a}s}},\ }\href {http://stacks.iop.org/1367-2630/17/i=5/a=053048}
  {\bibfield  {journal} {\bibinfo  {journal} {New J. Phys.}\ }\textbf {\bibinfo {volume}
  {17}},\ \bibinfo {pages} {053048} (\bibinfo {year} {2015})}\BibitemShut
  {NoStop}%
  \bibitem [{\citenamefont {Eisert}\ \emph {et~al.}(2016)\citenamefont
  {Eisert}, \citenamefont {Eisler},\ and\ \citenamefont
  {Zimbor{\'a}s, Zolt{\'a}n}}]{eisert2016entanglement}%
  \BibitemOpen
  \bibfield  {author} {\bibinfo {author} {\bibfnamefont {J.}\ \bibnamefont
  {Eisert}}, \bibinfo {author} {\bibfnamefont {V.}~\bibnamefont {Eisler}}, \
  and\ \bibinfo {author} {\bibfnamefont {Z.}~\bibnamefont {Zimbor{\'a}s}},\ }\href@noop
  {} {\  (\bibinfo {year} {2016})},\ \Eprint {http://arxiv.org/abs/1611.08007}
  {arXiv:1611.08007 [quant-ph]} \BibitemShut {NoStop}%
\bibitem [{\citenamefont {Botero}\ and\ \citenamefont
  {Reznik}(2004)}]{Alonso2004}%
  \BibitemOpen
  \bibfield  {author} {\bibinfo {author} {\bibfnamefont {A.}~\bibnamefont
  {Botero}}\ and\ \bibinfo {author} {\bibfnamefont {B.}~\bibnamefont
  {Reznik}},\ }\href {\doibase
  http://dx.doi.org/10.1016/j.physleta.2004.08.037} {\bibfield  {journal}
  {\bibinfo  {journal} {Phys. Lett. A}\ }\textbf {\bibinfo {volume} {331}},\
  \bibinfo {pages} {39 } (\bibinfo {year} {2004})}\BibitemShut {NoStop}%
\end{thebibliography}
\end{document}